%% file: main.tex
\documentclass[journal=jacsat,manuscript=article]{achemso}

\usepackage[version=3]{mhchem} 
\usepackage{amsmath}
\usepackage{amssymb}
\usepackage{pifont}
\usepackage{siunitx}
\usepackage{xcolor}
\usepackage{enumitem} 
\usepackage{tikz}
\usepackage{tabularx}
\usepackage{graphicx}
\usepackage{adjustbox}
\usepackage{multirow}
\usepackage{makecell}
\usepackage{subfiles} 
\author{Abin Varghese}
\affiliation[IITB]
{Department of Electrical Engineering, Indian Institute of Technology Bombay, Mumbai 400076, India}
\alsoaffiliation[Monash]{Department of Materials Science and Engineering, Monash University, Clayton, Victoria 3800, Australia}
\alsoaffiliation[IITBMonash]{IITB-Monash Research Academy, IIT Bombay, Mumbai 400076, India}
\author{Adityanarayan Pandey}
\affiliation[IITB]
{Department of Electrical Engineering, Indian Institute of Technology Bombay, Mumbai 400076, India}
\author{Pooja Sharma}
\affiliation[IITB]
{Department of Electrical Engineering, Indian Institute of Technology Bombay, Mumbai 400076, India}
\author{Yuefeng Yin}
\affiliation[Monash]{Department of Materials Science and Engineering, Monash University, Clayton, Victoria 3800, Australia}
\author{Nikhil V Medhekar}
\affiliation[Monash]{Department of Materials Science and Engineering, Monash University, Clayton, Victoria 3800, Australia}
\author{Saurabh Lodha}
\affiliation[IITB]
{Department of Electrical Engineering, Indian Institute of Technology Bombay, Mumbai 400076, India}
\email{slodha@ee.iitb.ac.in}

\title[An \textsf{achemso} demo]
  {Electrically Controlled Reversible Strain Modulation in \ce{MoS2} Field-effect Transistors via an Electro-mechanically Coupled Piezoelectric Thin Film}

\begin{document}
\maketitle

\subfile{manuscript}

\pagebreak

\renewcommand{\thefigure}{S\arabic{figure}}
\begin{center}
\begin{LARGE}  
\textsc{Supplementary Information}   
\end{LARGE}   
\end{center}
\subfile{SI}

\end{document}

%% file: manuscript.tex

\begin{abstract}
Strain can efficiently modulate the bandgap and carrier mobilities in two-dimensional (2D) materials. Conventional mechanical strain-application methodologies that rely on flexible, patterned or nano-indented substrates are severely limited by low thermal tolerance, lack of tunability and/or poor scalability. Here, we leverage the converse piezoelectric effect to electrically generate and control  strain transfer from a piezoelectric thin film to electro-mechanically coupled ultra-thin 2D \ce{MoS2}. Electrical bias polarity change across the piezo film tunes the nature of strain transferred to \ce{MoS2} from compressive ($\sim$0.23$\%$) to tensile ($\sim$0.14$\%$) as verified through peak shifts in Raman and photoluminescence spectroscopies and substantiated by density functional theory calculations. The device architecture, built on a silicon substrate, uniquely integrates an \ce{MoS2} field-effect transistor on top of a metal-piezoelectric-metal stack enabling strain modulation of transistor drain current ($130\times$), on/off current ratio ($150\times$), and mobility ($1.19\times$) with high precision, reversibility and resolution. Large, tunable tensile (1056) and compressive (-1498) strain gauge factors, easy electrical strain modulation, high thermal tolerance and substrate compatibility make this technique promising for integration with silicon-based CMOS and micro-electro-mechanical systems.

\end{abstract}

\section{Introduction}
2D materials, specifically transition metal dichalcogenides (TMDs) like \ce{MoS2}, demonstrate sizeable bandgaps, excellent electrostatic control, and remarkable light absorption\cite{thakar2020optoelectronic}. Their critical material and device properties such as bandgap,\cite{mak2010atomically, johari2012tuning} absorption coefficient,\cite{zhao2013evolution} carrier effective mass,\cite{scalise2014} electrical\cite{ghorbani2013strain} and thermal conductivities,\cite{zhang2015measurement} dielectric constant,\cite{kang2021local} and carrier mobility,\cite{kim2016thickness} exhibit a strong flake-thickness (number of layers) dependence.\cite{chaves2020bandgap} For example, the optical bandgap can be tailored from \SI{1.3}{eV} to \SI{1.9}{eV} as \ce{MoS2} is thinned from the few-layer regime to its monolayer\cite{mak2010atomically}. Hence, once a 2D material flake is exfoliated or grown at a specific thickness, its layer number-specific properties cannot be easily modulated. 
\\
Strain engineering, which relies on the application of mechanical strain to alter structural material parameters and hence their various mechanical, optical, thermal or electrical properties, has been successfully employed for conventional semiconductors such as silicon and germanium.\cite{thompson200490} In the case of 2D materials, their ultra-thin nature and high tensile strength can lead to a significant influence of strain on their electronic bandstructure\cite{johari2012tuning,scalise2014}. This can lead to a strong strain dependence of their photoluminescence peak positions, Raman modes, carrier mobilities, and electrical transport.\cite{peng2020strain} For example, the application of uniaxial tensile strain on monolayer \ce{MoS2} can lead to a 120 meV$\slash \%$ reduction in its optical bandgap.\cite{conley2013bandgap,shen2016strain} Modulation of carrier transport in strained monolayer \ce{MoS2}-based field-effect transistors (FETs) has also been demonstrated.\cite{datye2022,jaikissoon2022mobility, ng2022improving} 
\\
Several strain application strategies such as substrate doping, strain-incorporated substrate growth, flexible/bendable substrates, tensile/compressive capping layers, patterned substrates and atomic force miscroscopy (AFM)-based localised indentation, have been reported in literature. However, conventional techniques of introducing strain, i.e.\ by doping or strain-incorporated growth (e.g.\ Si$_{1-x}$Ge$_{x}$ source-drain growth in p-type Si FETs) are not viable for 2D materials since, being atomically thin, lattice-level alterations mostly lead to a deterioration of intrinsic properties.\cite{fox2013helium} The most explored method to apply strain on 2D flakes has been using flexible substrates such as poly-dimethylsiloxane (PDMS)\cite{yang2014lattice}, polyethylene terephthalate (PET)\cite{wu2014piezoelectricity}, polyimide (PI) \cite{li2020efficient}, polyvinyl alcohol (PVA)\cite{liu2014strain} or polyethylene naphthalate (PEN)\cite{datye2022}. Such flexible substrates cannot be easily integrated for applications involving high temperature processing that can affect their quality and rigidity. Moreover, device fabrication using organic polymer substrates could leave significant residues which can degrade their (opto)electronic performance.\cite{pak2020strain} Patterned or distorted substrates can also strain 2D materials, but without tunability or reversibility, similar to tensile/compressive capping layers.\cite{pena2021strain, jaikissoon2022mobility} AFM tips have been shown to locally strain TMDs, however, such techniques are unsuitable for device-level applications.\cite{manzeli2015piezoresistivity, quereda2017strain} Hence, a strain application strategy that can tune the properties of 2D materials in a controlled and reversible manner, and also that can be easily integrated with existing device fabrication process flows, is needed to harness their full potential for technological applications.
\\
Electrically controlled strain modulation in 2D materials-based devices is one such strategy that is absent in the strain application methods discussed above. Piezoelectric and electrostrictive materials demonstrate a coupling between their electrical and mechanical properties and are ubiquitous in micro-electro-mechanical systems (MEMS) and energy harvesters.\cite{li2014electrostrictive,safaei2019review} Sub-micron piezoelectric film-based mechanical actuation has been demonstrated with timescales in the order of nanoseconds at low actuation voltages.\cite{judy2009piezoelectric, smith2012pzt} In addition, piezoelectric thin films have shown low fatigue (loss of polarization and hence the elastic strain) due to repeated electrical-mechanical switching over multiple cycles ($10^4$) implying high endurance for device applications.\cite{jiang1993effects,lupascu2005fatigue} In conventional MEMS devices, strain transfer to a piezoelectric sensor or transducer results in a change in electrical resistance. Using the same principle, the converse piezoelectric effect can transfer strain from an electrically biased piezoelectric material to another material in close proximity. This strategy could be employed for an electrically controlled transfer of strain from a piezoelectric thin film to a 2D material.\cite{hui2013,chakraborty2020strain} Not only is it possible to control the piezoelectric thin film-based strain transfer with high resolution, precision and reversibility using electrical bias, but also to easily integrate it (i) with an FET that allows controlled strain-dependent tuning of electrical device parameters such as currents and carrier mobilities, and, (ii) with a conventional, rigid \ce{Si} substrate with high thermal tolerance and compatibility with existing CMOS and MEMS device technologies. 
\\
In this work, we report an all-electrical, highly controllable transfer of strain onto an atomically thin 2D material flake by conjugating it with a piezoelectric thin film on a Si substrate. The substantially large electric field induced distortions of the piezoelectric film, with large piezoelectric coefficients, can be efficiently transferred to an ultra-thin \ce{MoS2} flake via the converse piezoelectric effect. This precise electrical modulation of the magnitude and nature of strain in \ce{MoS2} is examined using Raman spectroscopy, yielding a shift in the $E'$ Raman mode by $0.26\pm 0.05$ \SI{}{cm^{-1}/V} for compressive and $-0.15\pm 0.02$ \SI{}{cm^{-1}/V} for tensile strains. Photoluminescence (PL) measurements and piezoresponse force microscopy (PFM) further reinforce the material-level strain transfer from the piezo film to the \ce{MoS2} flake. The device architecture offers a unique advantage of integrating a strain-tunable field-effect transistor, in which the two- and three-terminal electrical transport parameters of the ultra-thin \ce{MoS2} flake are readily modulated by the piezo biasing. Thus, we demonstrate significant strain-induced modulation of two-terminal \ce{MoS2} current ($130 \times$ across \SI{6}{V} of piezo bias). In addition, field-effect parameters such as the on/off current ratio ($7\times$), threshold voltage (shift of \SI{0.44}{V}) and mobility ($1.19\times$), as well as structural parameters such as Raman modes and the optical bandgap can be tailored  by the piezo biasing. The electrical bias controlled strain transfer feature of the device architecture not only offers easy, fast and precise strain tunability at high resolution, in comparison to conventional mechanical techniques, but it also (i) enables switching between tensile and compressive strains based on the bias polarity, and, (ii) unambiguously deconvolves strain versus field-effect components of electrical parameter change using different biasing schemes. We demonstrate large strain gauge factors for both compressive (-1498) and tensile (1056) strains indicating efficient sensing of piezo-induced strain in \ce{MoS2}. This demonstration of a device platform for electrically controlled strain transfer and modulation of 2D material-based transistor performance, on conventional silicon substrates, can enable the development of CMOS compatible, strain-tunable 2D materials-based MEMS circuits for sensing as well as energy harvesting applications. 

\section{Results and Discussion}
\subsection{Device Architecture}
A mixed-dimensional heterostructure device was designed for transferring electrically-induced strain in a piezoelectric thin film to the 2D material flake via the converse piezoelectric effect as demonstrated in Figure 1a. From bottom up, the device architecture consists of three major components, (i) a bottom metal electrode (M)-piezoelectric thin film (P)-top metal electrode (M) MPM stack on an \ce{Si}/\ce{SiO2} substrate, (ii) an intermediate dielectric layer over the top electrode that couples the bottom MPM stack to, (iii) a micro-mechanically exfoliated and source/drain metallized 2D material flake FET on top of the dielectric layer. A 3D schematic of the device configuration is shown in Figure 1b and the complete vertical cross-section as well as an optical microscope image of the as-fabricated device are shown in Figures 1c, d, respectively. 
\begin{figure}
  \includegraphics[width=\textwidth,height=\textheight,keepaspectratio]{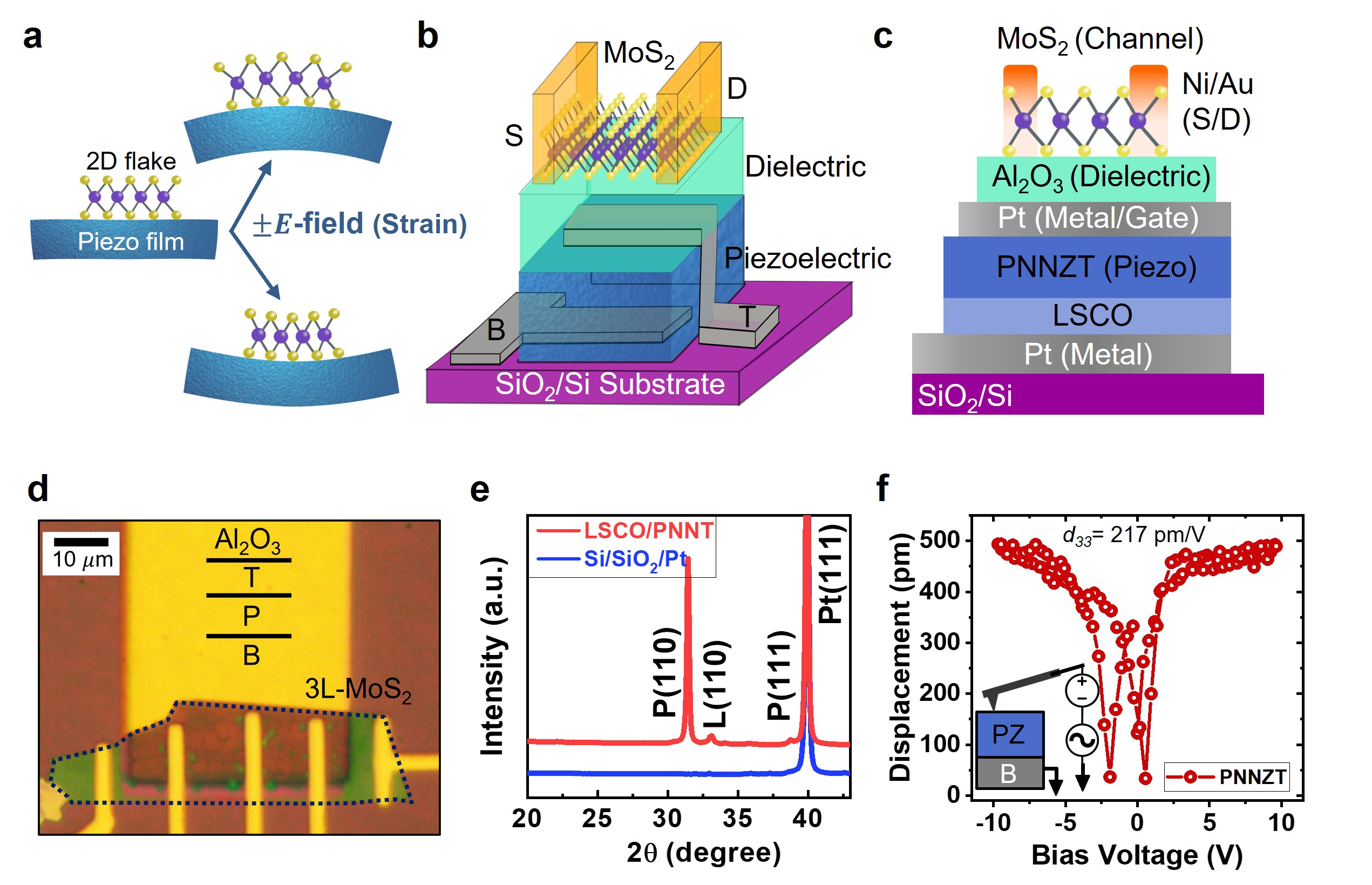}
   \caption{\textbf{Device architecture.} \textbf{a} Pictorial representation of the converse piezoelectric effect based all-electrical strain transfer from a piezoelectric thin film to the 2D material. \textbf{b} Schematic of the \ce{Si} substrate-based piezoelectric thin film-2D material electro-mechanical device, and, \textbf{c} a vertical cross-section of the device showing the metal-piezoelectric-metal (MPM) stack at the bottom and the \ce{MoS2} field-effect transistor (FET) with source (S)-drain (D) contacts on top, coupled through the intermediate \ce{Al2O3} dielectric layer. \textbf{d} Optical microscope image of the as-fabricated device with different regions marked, T and B are top and bottom electrodes of the MPM stack and P represents the sandwiched piezoelectric thin film. \textbf{e} XRD plots of LSCO\slash PNNZT film and platinized Si substrate for reference, and, \textbf{f} shows the strain hysteresis loop of the PNNZT film obtained by piezoresponse force microscopy.}
\label{sch:Figure1}   
\end{figure}
\\
The piezoelectric thin film, here a complex perovskite oxide \ce{0.50Pb(Ni1/3Nb2/3)O3} $-$ \ce{0.35PbTiO3} $-$ \ce{0.15PbZrO3} (PNNZT), was grown using pulsed laser deposition (PLD) at high temperature (\SI{750}{\degreeCelsius}). Single-phase targets of PNNZT were used for the deposition to produce thin films with thickness in the range of \SI{350}{} $-$ \SI{500}{nm}, and their quality was ascertained through x-ray diffraction (XRD) measurements. It is crucial to obtain the pure perovskite phase, devoid of any pyrochlore phase, for achieving significant piezoelectric parameters. For this, the PNNZT thin film was grown on a buffer layer of La$_{0.5}$Sr$_{0.5}$CoO$_3$ (LSCO) or \ce{SrTiO3} (STO), also deposited using PLD, on a Pt/Ti/\ce{SiO2}/Si substrate. The XRD patterns of LSCO/PNNZT film are shown in Figure 1e. Bragg reflections from the (110) and (111) planes were obtained at 2$\theta$ = \SI{31.44}{\degree} and \SI{38.74}{\degree} respectively, indicating the formation of the perovskite phase.\cite{ade2018} The XRD peak at \SI{33.12}{\degree} corresponds to the (110) plane of LSCO. 
\\
The principal metrics that relate the mechanical strain to an applied electric field  of a material are the piezoelectric voltage coefficients, $d_{ij}$. The first index $i$ is the direction of application of electric field and the second index $j$ corresponds to the direction along which the piezoelectric effect is evaluated. Typically, 1 and 2 are the in-plane directions (along the surface) and 3 is the out-of plane direction. The coefficients of importance are $d_{33}$ (both electric field and surface displacement along the $z$-direction) and $d_{31}$ (the in-plane distortion resulting from an electric field perpendicular to the surface). PNNZT was chosen for its reported sizeable piezoelectric voltage coefficient ($d_{33}$) $\sim$ \SI{250}{pm/V}, which is higher than commonly used piezoelectric films of \ce{Pb(Zr, Ti)O3} (PZT, \SI{100}{pm/V})\cite{trolier2004PZT} or \ce{BaTiO3} (BTO, \SI{20}{pm/V})\cite{kelley2020BTO}. Further details of the growth process are available in sections S1 and S2 of Supplementary Information.
\\
The electro-mechanical response of the PNNZT films was evaluated using the PFM technique (Figure 1f). In this, the film's surface was scanned using a conductive tip, biased with an AC voltage superimposed on a DC bias, which can sense the local electric field-dependent deformations on the surface via the converse piezoelectric effect. This can also capture the local phase switching of the piezoelectric domains in the film. Upon straining, the critical component of the piezoelectric tensor ($d_{33}$) couples directly to the vertical (out-of-plane) deflection of the PFM cantilever. The typical extracted values of $d_{33}$ were in the range of \SI{150}{} to \SI{250}{pm/V} (Figure S8, Supplementary Information). The MPM stack also allows the determination of ferroelectric parameters of the film, like the large remnant polarization of 29.5 $\mu$C/cm$^2$.
\\
For the device implementation, Ti(\SI{5}{nm})/Pt(\SI{50}{nm}) bottom electrodes (B) were lithographically patterned, and LSCO/PNNZT films were PLD-grown at elevated temperature, followed by the electron-beam patterning and deposition of a Ti(\SI{3}{nm})/Pt(\SI{20}{nm}) top electrode (T) to complete the formation of the MPM stack. A thin film ($\sim$ \SI{15}{nm}) of \ce{Al2O3} was deposited using atomic layer deposition (ALD) on top of the MPM stack, and the resulting vertical structure forms the active strain transfer region. Thereafter, an ultra-thin flake of exfoliated \ce{MoS2} was transferred onto the active region, and source and drain electrodes were formed using electron-beam patterning and a sputter-deposited Ni(\SI{20}{nm})/Au(\SI{80}{nm}) metal stack. An atomic force microscope scan of the \ce{MoS2} flake showing its thickness ($\sim$\SI{2.4}{nm}) is available in S3 of Supplementary Information (Figure S5). 
\\
\subsection{Nature of Strain Transfer}
\begin{figure}
\includegraphics[width=\textwidth,height=\textheight,keepaspectratio]{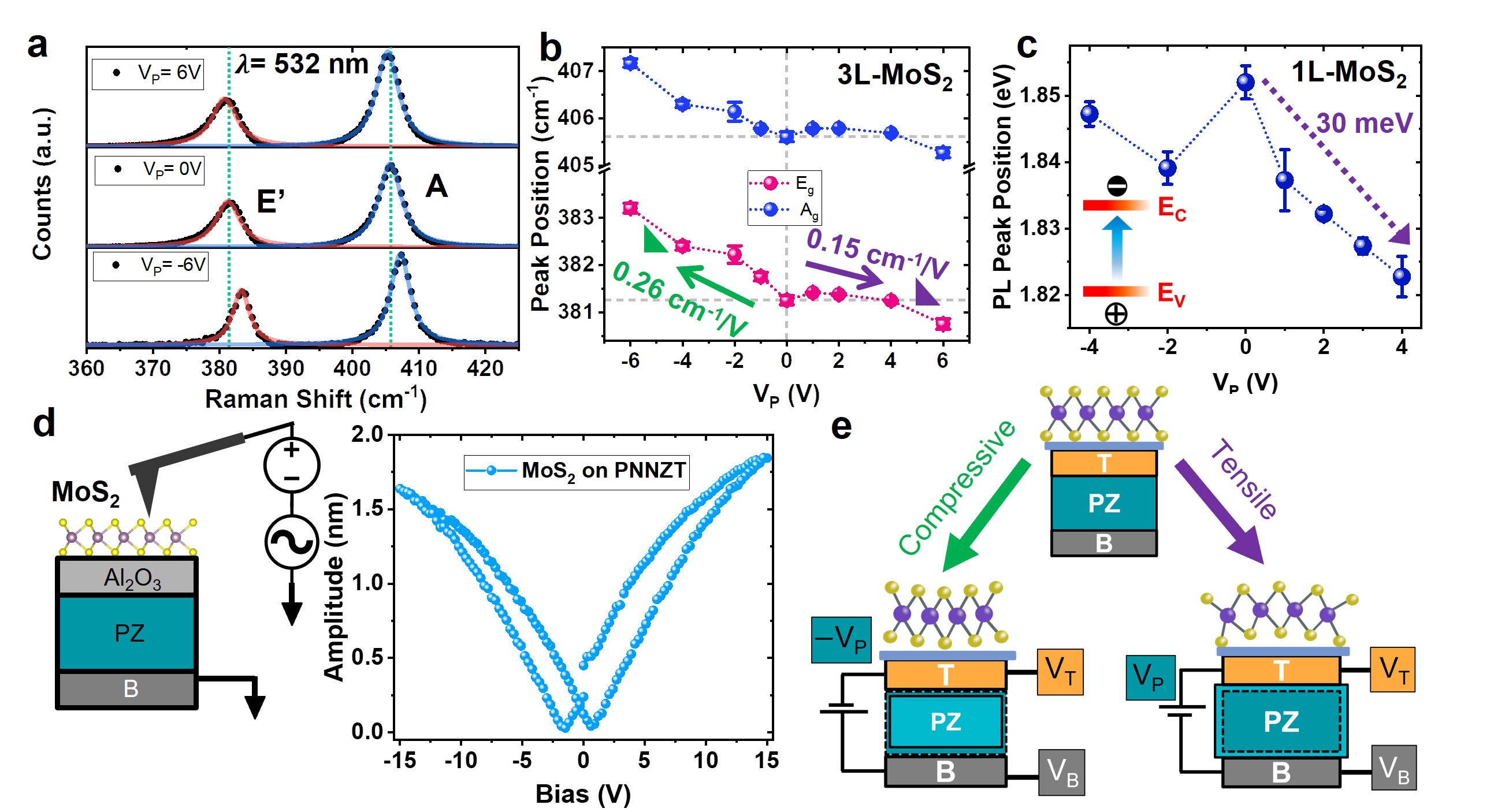}
   \caption{\textbf{Physical evidence of strain transfer.} \textbf{a} Raman spectra (using $\lambda$= \SI{532}{nm} laser) of \ce{MoS2} under unstrained ($V_P$= \SI{0}{V}), and strained ($V_P$= \SI{-6}{V} and $V_P$= \SI{6}{V})  conditions. The two characteristic modes have been fit using Lorentzian functions. \textbf{b} Raman shifts of $E'$ and $A$ modes with piezo biasing, showing a redshift for positive $V_P$ and blueshift for negative $V_P$ corresponding to tensile and compressive strains, respectively. \textbf{c} Photoluminescence peak positions of monolayer \ce{MoS2} for varying $V_P$. The $A$ excitonic peak shifts by \SI{-30}{meV} when $V_P$ of \SI{4}{V} is applied. \textbf{d} Pictorial depiction of PFM imaging on an \ce{MoS2} flake on top of the \ce{Al2O3}/piezoelectric stack, and, PFM strain loop demonstrating the bias-dependent localised displacement of the \ce{MoS2} flake in response to the vertical electric field. \textbf{e} Schematic representation of the nature of strain transferred to \ce{MoS2} as determined by the polarity of $V_P$. The unstrained physical dimensions are denoted by the dotted lines and the $V_P$-induced strained dimensions are represented by solid lines.}
\label{sch:Figure2}   
\end{figure}
Bias voltages applied to the top ($V_T$) and bottom ($V_B$) electrodes result in an out-of-plane electric field, $\vec{E}$= $\frac{V_B-V_T}{d}\hat{z}$=$\frac{V_P}{d}\hat{z}$, where $d$ is the thickness and $V_P= V_B-V_T$ is the potential difference across the piezoelectric film, ranging from 0 ($V_P$= \SI{0}{kV}) to about $\pm$\SI{150   }{kV/cm} ($V_P$= $\pm$ \SI{6}{V}). If the polarization of the ferroelectric domains in PNNZT is aligned with the direction of the electric field, out-of-plane as well as in-plane deformations can be produced depending on the correlated $d_{33}$ and $d_{31}$ of PNNZT. The nature of strain (compressive/tensile) and its magnitude can be controlled by the polarity and magnitude of the piezo voltage respectively. 
\\
The resulting strain transferred onto the \ce{MoS2} flake was examined by bias-dependent Raman spectroscopy using a \SI{532}{nm} laser. The two prominent Raman active modes of \ce{MoS2}, $E'$ and $A_1'$, are sensitive to the application of strain that distorts its hexagonal Brillouin zone. The in-plane $E'$ mode is more sensitive to strain in comparison to the out-of-plane $A_1'$ mode. In Figure 2a, $V_P$-dependent shifts in the characteristic phonon modes of a three-layer (3L) \ce{MoS2} flake are shown. As $V_P$ is increased in the negative direction, the $E'$ mode shifts to higher frequencies at a rate of $0.26\pm 0.05$ \SI{}{cm^{-1}/V}. On the other hand, for increasing positive $V_P$, $E'$ shows a redshift at $0.15\pm 0.02$ \SI{}{cm^{-1}/V}. First-principles based calculations have shown a stiffening of phonon modes with compressive strain and a relative softening with tensile strain\cite{scalise2014}. This trend in the shift in $E'$ mode with compressive/tensile strain is consistent with trends seen in \ce{MoS2} flakes strained by other techniques\cite{datye2022,hui2013}. 
\\
In addition, for the case of monolayer (1L) \ce{MoS2}, by monitoring the direction of shift of the $A$ excitonic peak (\textbf{K} $-$ \textbf{K} transition) energies in the bias-dependent PL spectra, the nature of strain transfer with applied $V_P$ can be confirmed. The PL peak shift with piezo biasing in a strained 1L \ce{MoS2} flake is depicted in Figure 2c. For increasing positive $V_P$, the $A$ peak shows a significant shift towards lower energies, whereas, for higher negative $V_P$ values, the PL peak shifts are smaller. This observation, as also seen in literature\cite{pak2017strain},  can be directly correlated with the change in the bandstructure of \ce{MoS2} with strain as shown in our density functional theory based first-principles calculations in S4 of Supplementary Information. The decrease in bandgap of monolayer \ce{MoS2} is shown to be much steeper for tensile strain (both uniaxial and biaxial) in comparison to compressive strain.\cite{lopez2016} Hence, the direction of peak shifts in the bias-dependent Raman and PL spectra indicates the nature of strain transfer to the \ce{MoS2} flake by the application of the piezo voltage $-$ a positive $V_P$ results in tensile straining while negative $V_P$ leads to compressive straining of the \ce{MoS2} flake. 
\\
Next, to probe the transfer of strain from the piezoelectric layer to the \ce{MoS2} flake, we performed PFM analysis of the \ce{MoS2}/\ce{Al2O3}/PNNZT stack. With the application of a DC bias through the PFM tip, the piezoelectric polarization-induced displacement of PNNZT that gets transmitted to \ce{MoS2} can be captured. The obtained strain-loop in Figure 2d shows that the electric field-induced deformation of the PNNZT film can be sensed at the 2D material surface. The hysteresis strain-loop of an \ce{MoS2} flake on non-piezoelectric substrates like \ce{Si}\slash \ce{SiO2} exhibits no evidence of straining as can be seen in S5 of Supplementary Information. A schematic summarizing the nature of strain transfer with piezo biasing is shown in Figure 2e. 
\\
\subsection{Electrical Characteristics}
\begin{figure}
  \includegraphics[width=\textwidth,height=\textheight,keepaspectratio]{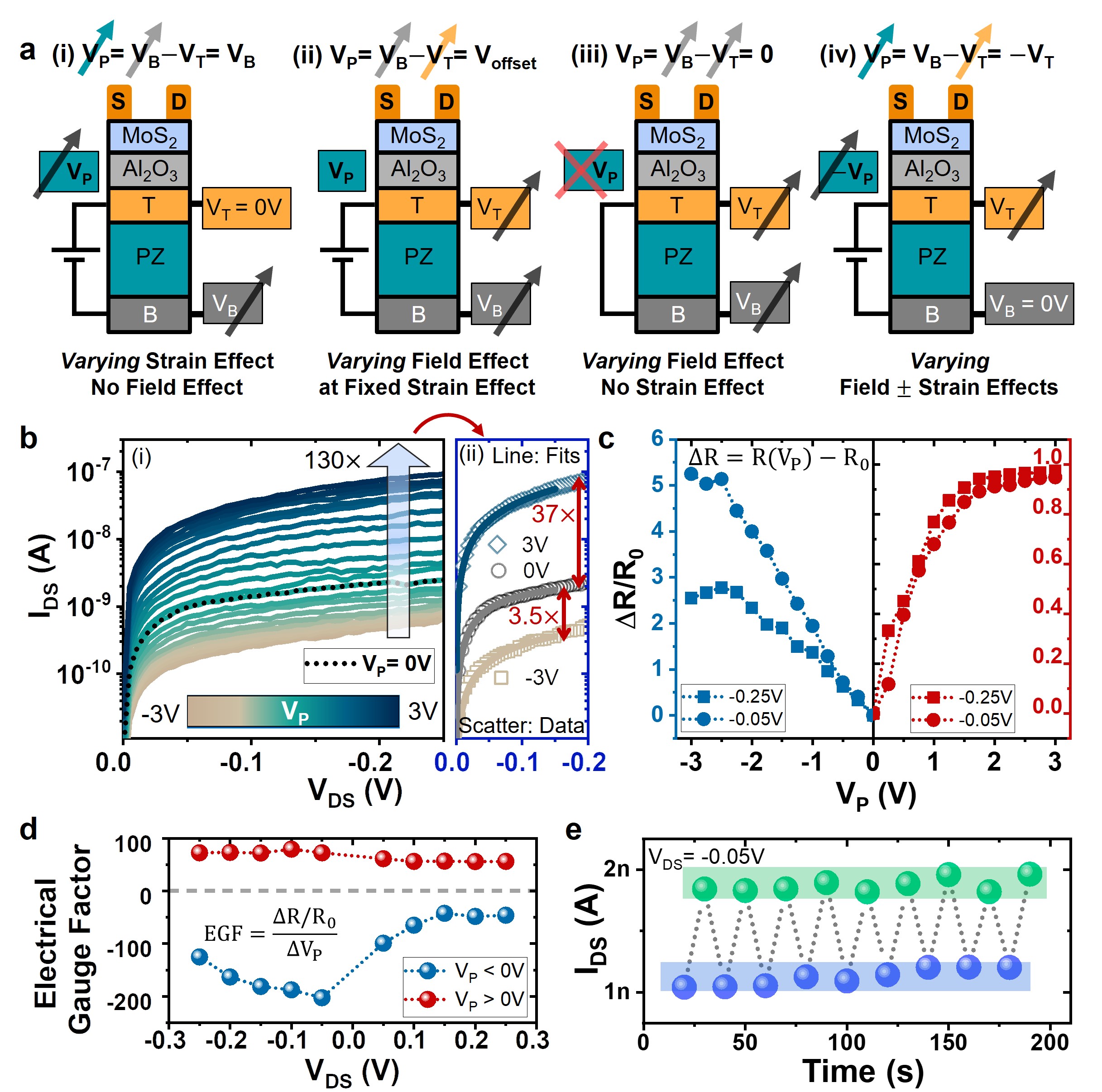}
   \caption{\textbf{Strain modulated two-probe electrical transport.} \textbf{a} Various biasing conditions to deconvolve the interplay between strain-effect and field-effect in the coupled MPM-\ce{MoS2} FET device. \textbf{b} (i) Two-terminal $I_{DS}$-$V_{DS}$ characteristics (biasing scheme (i) in (a)) showing the modulation of drain current with strain applied by varying $V_P$ at zero $V_T$ (no field-effect). The output trace for the unstrained case, $V_P$= \SI{0}{V}, is shown by the black dotted line. (ii) Analytical current model based fit lines of the $V_P$= -3, 0, 3 \SI{}{V} traces overlaid on the scatter data points. A $37\times$ enhancement of the drain current at \SI{3}{V} and $3.5\times$ reduction at \SI{-3}{V} is obtained with the piezo biasing. \textbf{c} Plot of the normalised change in resistance with $V_P$ for compressive (left plot, in blue) and tensile (right plot, in red) strains extracted for two $V_{DS}$ values. The electrical gauge factor in \textbf{d} is calculated from the slopes of the respective $V_{DS}$ plots in \textbf{c}. \textbf{e} Switching between the unstrained ($V_P$= \SI{0}{V}) and strained ($V_P$= \SI{1}{V}) drain current values over multiple cycles.}
\label{sch:Figure3}   
\end{figure}
To understand the DC electrical characteristics of the electro-mechanically coupled piezoelectric thin film MPM and \ce{MoS2} FET devices, we first take a closer look at the device architecture. The top electrode of the MPM stack is capacitively coupled to the \ce{MoS2} channel through the \ce{Al2O3} dielectric. Hence, it can also act as the gate electrode of an FET with an n-type \ce{MoS2} channel with source and drain electrodes on top. The series connection of the MPM stack and the FET through the shared top (gate) electrode gives rise to the possibility of studying strain and field-effect driven transport in \ce{MoS2} individually and in combination, through different biasing schemes, as depicted in Figure 3a. Specifically, measuring source-drain current while (i) grounding the top electrode ($V_T$= \SI{0}{V}) and sweeping the bottom electrode bias (varying $V_B$) leads to strain-dependent transport, while (ii) sweeping $V_T$ and $V_B$ together with a fixed $V_P$ offset ((iii), zero offset) voltage results in field-effect transport under a fixed strain value ((iii), just the field-effect transport with no strain-effect), and (iv) grounding $V_B$ and sweeping $V_T$ gives the combined effect of  simultaneously varying field-effect and strain on \ce{MoS2} transport.
\\
Of these four biasing schemes, we look at the first one investigating strain-dependent two terminal transport in \ce{MoS2} as depicted in Figure 3b. The top electrode is kept fixed at $V_T$= \SI{0}{V} and the current-voltage ($I_{DS}-V_{DS}$) curves are acquired at different bias voltages on the bottom electrode, $V_B$. As described earlier, in this configuration, $V_P= V_B$ and any change in carrier concentration in the \ce{MoS2} channel due to field-effect will be negligible since the gate voltage ($V_T$) is zero. As $V_P$ is increased from \SI{-3}{V} to \SI{3}{V}, the two-probe current through the device increases, as shown in Figure 3b. Increasing $V_P$ in the positive direction (tensile strain) leads to an increase in the drain current, whereas decreasing $V_P$ in the negative direction (compressive strain) leads to a reduction in current. The drain current trace for $V_T = V_B =$ \SI{0}{V} is highlighted to clearly distinguish the change in current with $V_P$. A \SI{6}{V} change in piezo biasing can modulate the drain current by nearly 130$\times$. 
\\
From the two-probe output characteristics, the change in resistance of the channel with piezo voltage can be determined at different values of the drain bias, $V_{DS}$. The resistance at $V_P$= \SI{0}{V} is noted as $R_0$ and the change in resistance with $V_P$ is calculated as $\Delta R = R(V_P) - R_0$ for a set value of $V_{DS}$. The normalised change in resistance ($\frac{\Delta R}{R_0}$) with piezo voltage is calculated in Figure 3c for both compressive and tensile strains at $V_{DS} =$ \SI{-0.25}{} and \SI{-0.05}{V}. Linear fits to the plot in Figure 3c provide a measure of the sensitivity of the material as a strain sensor. This metric, gauge factor ($GF$), is conventionally calculated as the change in normalised resistance with strain, $GF = \frac{\Delta R}{R_0}/\Delta \epsilon$, where $\epsilon$ is the applied strain. Since our device is electrically strained, the gauge factor can be defined in terms of the piezo voltage. Thus, the electrical gauge factor ($EGF= \frac{\Delta R}{R_0}/\Delta V_P$) can account for a change in the resistance of the channel with the application of strain through the piezo biasing. $EGF$ reaches a maximum value of -202 V$^{-1}$ for negative $V_P$ and a maximum value of 80 V$^{-1}$ for positive $V_P$. By calibrating the degree of piezo voltage dependent Raman shifts in our experiments with mechanical strain-based reports in literature, an equivalent strain $GF$ can be found out for our devices, which will be discussed later in the benchmarking section of the manuscript. Further, output characteristics of two additional devices showing good device-to-device repeatability through significant modulation of the drain current with piezo biasing are available in S6 of Supplementary Information.
\\
For a set value of $V_P$ (strain), the device shows good repeatability in current values across several cycles of the output characteristics. For ten back-to-back output sweeps (from $V_{DS}$= \SI{0}{V} to $V_{DS}$= \SI{-0.2}{V}), the drain current is $4.36\pm 0.32$ nA for the unstrained ($V_P$= \SI{0}{V}) case, while it increases to $7.57\pm 0.51$ nA at $V_P$= \SI{1}{V} tensile strain at a drain bias of \SI{-0.2}{V}; both cases depicting low standard deviation in their current values. A plot highlighting the measured error in the drain current values over the multiple sweeps is shown in S7, Supplementary Information. Further, to evaluate the time-dependent stability and current resolution of the drain current with strain, the device was switched between the unstrained and strained states over multiple cycles and the drain current levels were monitored. A clear distinction in the drain current values can be seen between the unstrained ($V_P$= \SI{0}{V}) and strained ($V_P$= \SI{1}{V}) biasing conditions as shown in Figure 3e.
\\
Next, to evaluate the field-effect performance of the \ce{MoS2} transistor under different fixed values of strain, we employ the biasing schemes shown in Figure 3a (ii) and (iii), and also in the inset in Figure 4a. In both cases, the top and bottom electrodes were connected and swept together but with a fixed offset ($V_B = V_T + V_{offset}$), where $V_{offset} = V_P \neq 0$ in case (ii) and $V_{offset} = V_P$ = \SI{0}{V} in case (iii). $V_{offset}$ ensures a fixed electric field (strain) between the top and bottom electrodes, while  $V_T$ and $V_B$ are swept together varying the gate field for the \ce{MoS2} channel on top. $V_{offset}$= \SI{0}{V} implies zero electric field (strain) for the MPM stack and gives us the unstrained, control, \ce{MoS2} transistor performance. Transfer characteristics ($I_{DS}-V_{T}$) of the \ce{MoS2} transistor, in Figure 4a, show a clear modulation of the drain current values with the offset voltage. Significant change in drain current values below the threshold voltage ($V_{Th}$) is shown in Figure 4b. Interestingly, Figure 4b shows that the drain current values for zero offset voltage before, during, and at the end of several $I_{DS}-V_T$ sweeps for different offset voltages (strain values) are nearly identical, highlighting the reversibility of the strain modulation. The corresponding $I_{DS}-V_T$ traces shown in S8 of Supplementary Information reinforce the recoverable (reversible) nature of the transfer curves with the offset biasing.
\begin{figure}
  \includegraphics[width=\textwidth,height=\textheight,keepaspectratio]{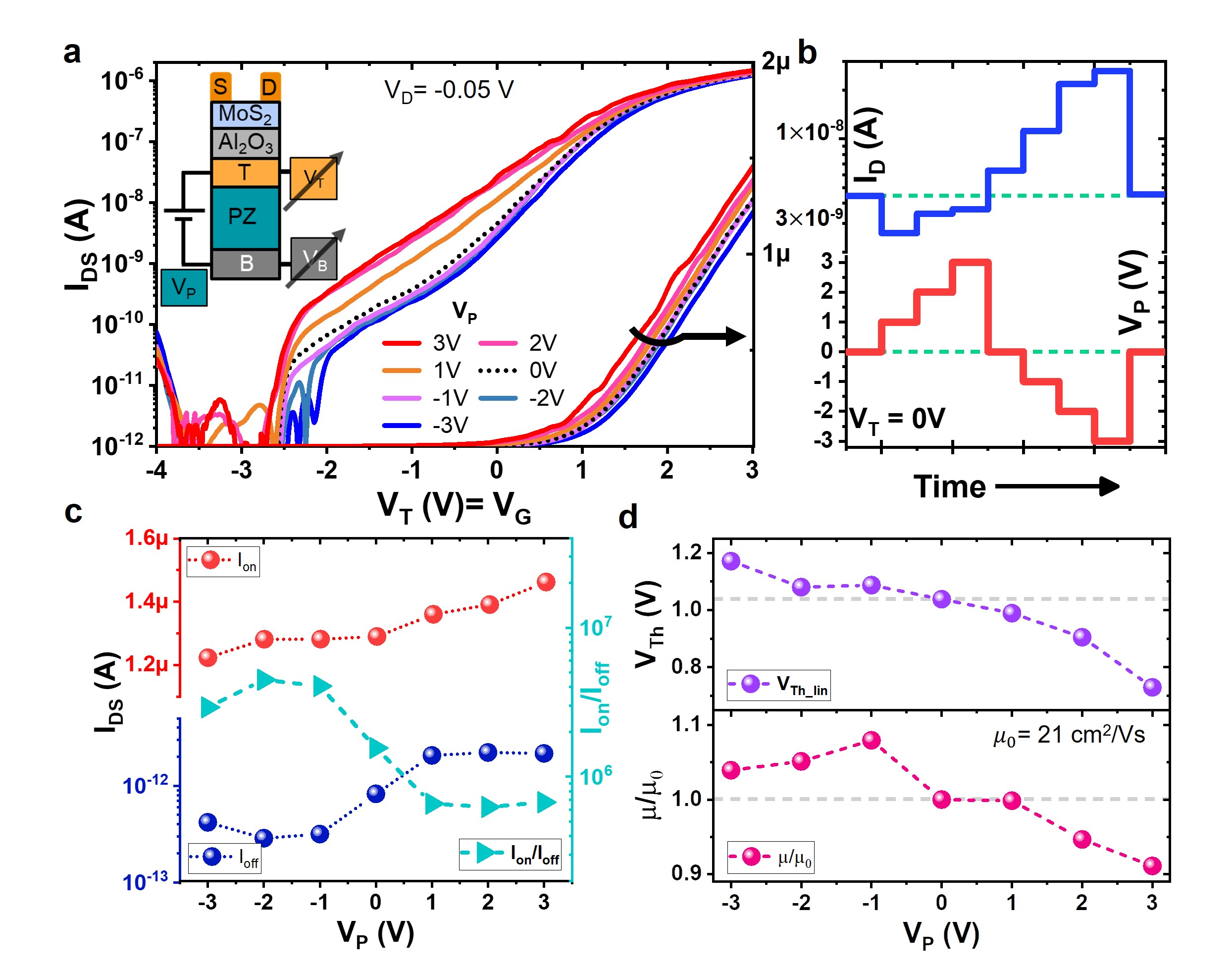}
   \caption{\textbf{Strain-tunable field-effect transistor.} \textbf{a} Transfer characteristics ($I_{DS}$-$V_{T}$ (gate bias)) of the \ce{MoS2} transistor at different values of piezo (offset) bias (strain). The biasing scheme displayed in the inset, also shown in Figure 3(a) (ii), was implemented for the measurements. The top and bottom electrodes were tied together and swept with different fixed offset biases, where the offset bias is essentially the piezo voltage $V_{P}$ that fixes the strain value for a given $I_{DS}$-$V_{T}$ sweep. \textbf{b} Dynamic evolution of the drain current at $V_{T}=\SI{0}{V}$, just below threshold voltage, when $V_P$ is increased from \SI{0}{V} to \SI{3}{V}, then decreased to \SI{-3}{V} and finally set at \SI{0}{V} indicating good reversibility of the strain modulation. \textbf{c} and \textbf{d} Modulation of the extracted FET parameters with $V_P$. Both $I_{on}$ and $I_{off}$ show an increasing trend with increasing $V_P$ (\SI{-3}{V} to \SI{3}{V}), however, the modulation of $I_{off}$ ($5.18\times$) is substantially higher than $I_{on}$ ($1.2\times$). Threshold voltage shifts to the left (negative direction) with increasing piezo voltage and the mobility can be tuned by 1.19$\times$ with $V_P$. In short, all FET parameters, $I_{on}$, $I_{off}$ , $V_{Th}$, and $\mu$, can be increased or decreased, in the strained-FET, with respect to the $V_P$= \SI{0}{V} FET, thereby highlighting strain as an additional knob to modulate field-effect parameters.}
\label{sch:Figure4}   
\end{figure}
\\
The FET parameters can be tuned substantially by the piezo biasing as shown in Figures 4c and d. Unlike a conventional FET, the on-current ($I_{on}$), the off-current ($I_{off}$) and the corresponding on-off ratio ($I_{on}/I_{off}$) can be modulated by nearly $1.2\times$, $5.18\times$ and $150\times$ respectively across a \SI{6}{V} piezo voltage range. The $V_{Th}$ for each transfer characteristic was obtained from its linear extrapolation at the maximum transconductance point, $g_m= dI_{DS}/dV_G$. $V_{Th}$ shifts to lower values with increasing $V_P$ which leads to an increase in $I_{on}$. Electron mobility values were extracted at $V_{Th}$ using $\mu = g_m\frac{L}{WC_GV_D}$, where $L$, $W$ are the length and width of the channel and $C_G$ is the capacitance of the \ce{Al2O3} gate dielectric (\SI{301}{nF/cm^2}). The unstrained mobility value of \SI{21}{cm^2/Vs}, at an offset ($V_P$) of \SI{0}{V}, can be tuned from $1.1\times$ to $0.9\times$ ($1.19\times$ overall) using piezo bias as shown in Figure 4d. In contrast to a recent report\cite{datye2022} on mobility enhancement with mechanically applied tensile strain in monolayer \ce{MoS2}, we observe a slight decrease in $\mu$ for positive $V_P$ (tensile strain) and an increase at negative $V_P$ (compressive strain). 
\\
We note that, for increasing positive $V_P$, the tensile strain leads to an increase in the on-current as reported previously. While $I_{on}$ increases by $1.2\times$, $I_{off}$ increases by $5.18\times$, which leads to a flattening of the slopes of the transfer curves and consequently a reduction in mobility. In our analysis, a crucial assumption is that the dielectric constant of \ce{Al2O3} is unchanged with strain. Thus, we can conclude that electrically applied strain can tailor the field-effect properties of the \ce{MoS2} flake. Computational studies have also predicted an improvement in the on-current of \ce{MoS2}-based transistors with tensile strain.\cite{sengupta2013} Further, on switching from compressive to tensile strain on \ce{MoS2} in a mechanical strain transfer system, the $V_{Th}$ shows a left-shift and the $I_{on}$ increases,\cite{lee2019ultrahigh} which is very similar to our observation. Thus, the electrical piezo-induced strain transfer methodology can reproduce the effect of mechanical straining techniques on 2D transistor performance, however, it enables a more precise control of the strain transfer with significantly improved resolution.
\\
Finally we evaluate biasing schemes shown in Figure 3a (iii) and (iv) depicting the impact of just varying the field-effect and combining it with a varying strain-effect. The two-probe output characteristics acquired using biasing scheme (iii) from Figure 3a, with no strain-effect but just the field-effect, show significant modulation of the drain current ($7.8\times10^4$ times) of the \ce{MoS2} channel (Supplementary Information, S9). In the biasing scheme (iv), considering that $V_{Th}$ of the \ce{MoS2} FET is around \SI{1}{V}, the \ce{MoS2} channel can be switched from the $off$ state ($V_T < V_{Th}$) to the $on$ state ($V_T > V_{Th}$) with the top electrode bias changing from \SI{-2}{V} to \SI{2}{V}. $V_B$ set at \SI{0}{V} leads to a nature of strain opposite to that observed in the biasing condition (i) (Figure 3b) since the direction of the electric field is now reversed. Hence, the drain current values can be compared for conditions (iii) and (iv) since both cases have non-zero biasing on the top electrode (Supplementary Information, Figure S18a, b). For the configuration (iv), firstly, under high positive $V_T$, the $I_{DS}$ values are similar to those obtained for case (iii) (red curves in Figure S13a, b) implying that the field-effect dominates over the compressive strain (since $V_B$=\SI{0}{V} and $V_B-V_T < 0$) that comes from the MPM stack. Secondly, it can be seen that for negative $V_T < V_{Th}$, when the transistor is in a high resistance state, its drain current values are higher than the $off$ currents in (iii) (compare blue and violet curves). This is because, the channel being in the $off$ state is not flooded by gate field-induced charge carriers, hence, tensile strain ($V_B$= \SI{0}{V} and $V_B-V_T > 0$) from the MPM stack dominates over the field-effect. As seen in Figure 3b, tensile strain can result in an increased drain current. Thus, this biasing scheme enables a strain-effect-based modulation of field-effect limited drain current ($1.4\times10^3$ times).
\\
To understand the strain-based modulation of electrical and optical characteristics we now describe the modulation of \ce{MoS2} bandstructure and its effect on current transport and PL measurements. We have performed first-principles based density functional theory calculations to understand the effect of compressive and tensile strain on the properties of \ce{MoS2}. To account for the mixed nature of strain that could be transferred from the piezoelectric layer to \ce{MoS2}, both uniaxial and biaxial strains were employed on a unit cell of 3L-\ce{MoS2}. For the unstrained case (inset in Figure 5a), there are two prominent valleys in the lowest energy conduction band, one at \textbf{K} and the other at \textbf{Q'} (along \textbf{K} $-$ \textbf{\boldmath$\Gamma$}) separated by \SI{92}{meV}. The valence band maximum is at \textbf{\boldmath$\Gamma$}, hence for the unstrained case, the bandgap is along \textbf{\boldmath$\Gamma$} $\rightarrow$ \textbf{Q'}. It should be noted that both the conduction band valleys will be sensitive to the nature and magnitude of applied strain. Further details of the strain-application methodology and the evolution of \ce{MoS2} bandstructure with strain are available in S4 of Supplementary Information. 
\\
Under the influence of in-plane tensile strain values, uniaxial or biaxial, the bandgap is along \textbf{\boldmath$\Gamma$} $\rightarrow$ \textbf{K} since the decrease in energy of the conduction band minimum valley at \textbf{K} point is much more significant. A steady decrease in the bandgap is observed for both cases, but biaxial strain leads to a steeper change of -76 meV/$\%$ as opposed to -19 meV/$\%$ for uniaxial strain. Next, for the case of increasing compressive strain values, the bandgaps remain along \textbf{\boldmath$\Gamma$} $\rightarrow$ \textbf{Q'} and show only a slight increase. These calculated changes in the bandgap values can be directly correlated to our strain-dependent experimental photoluminesence peak positions in Figure 2c. For example, a redshift shift of \SI{30}{meV} is observed in the PL peak position at $V_P$= \SI{4}{V}.
\begin{figure}
  \includegraphics[width=\textwidth,height=\textheight,keepaspectratio]{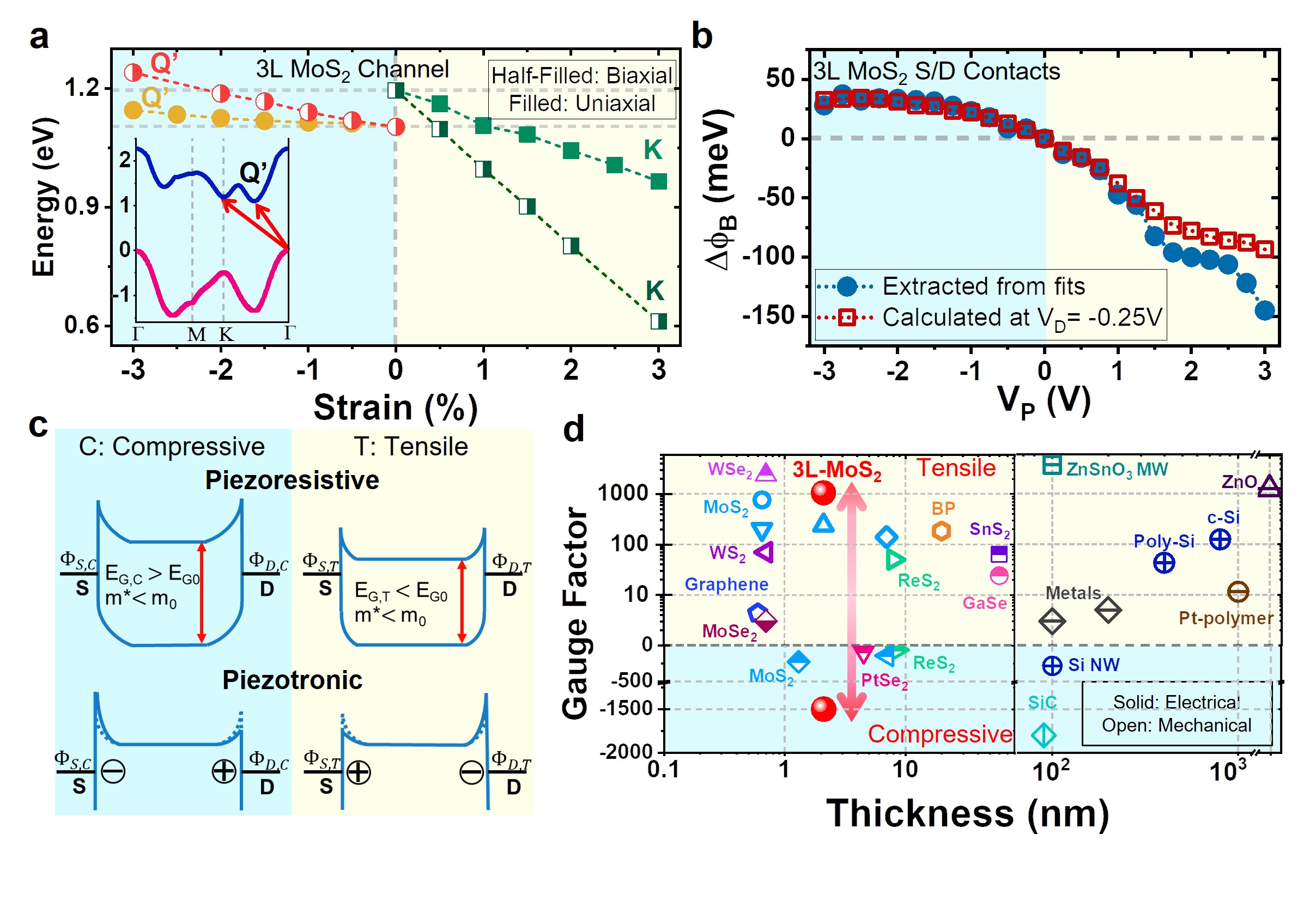}
   \caption{\textbf{Effect of strain on \ce{MoS2} bandstructure and contacts. Performance benchmarking.} \textbf{a} Calculated bandgap of 3L-\ce{MoS2} from first-principles. The electronic bandstructure for unstrained 3L-\ce{MoS2} is shown in the inset of \textbf{a}. The two lowest energy transitions are marked for \textbf{K} and \textbf{Q'} valleys. For both uniaxial and biaxial tensile strains, the bandgap along \textbf{\boldmath$\Gamma$} $\rightarrow$ \textbf{K}, shows a significant decrease, while for compressive straining of the unit cell, the bandgap occurs along \textbf{\boldmath$\Gamma$} $\rightarrow$ \textbf{Q'} and shows a comparatively smaller increase. \textbf{b} Change in Schottky barrier height with piezo bias obtained from, (i) extracted $V_P$-dependent Schottky barrier height values from the fits to $I_{DS}-V_{DS}$ characteristics in Figure 3b, and (ii) calculated using equation 2.\textbf{c} Representative band diagrams depicting the changes in bandgap and electron effective mass with compressive and tensile strains due to the piezoresistive effect and the Schottky barrier modulation due to accumulation of strain-induced polarization charges through the piezotronic effect. \textbf{d} Benchmarking of the gauge factor of our device with 2D materials and other quasi-2D and bulk materials reported in literature. The all-electrical tuning of strain along with ability to achieve both compressive and tensile strain response is highlighted. In all plots, regions highlighted by yellow and light blue colors refer to tensile and compressive strains, respectively.}
\label{sch:Figure5}   
\end{figure}
\\
The decrease in bandgap with tensile strain is predominantly due to the lowering of the conduction band minimum which can reduce the Schottky barrier height for electrons in the n-type \ce{MoS2} device. Thus, the change in drain current with strain can be effectively due to this piezoresistive effect (Figure 5c). In addition, we have also calculated the strain-tuned effective masses of electrons ($m^*_e$) around the conduction band minima at the \textbf{K} and  \textbf{Q'} valleys as shown in Figure S13, Supplementary Information. Under tensile strain, $m^*_e$ at the \textbf{K} valley shows a sharp reduction. Interestingly, for compressive strain, $m^*_e$ at the conduction band minimum (now around \textbf{Q'}) also follows a decreasing trend. Hence, under both tensile and compressive strains, the effective masses of electrons at their respective band minima decreases. The sharper decrease with increasing tensile strain could also lead to improved drain currents as observed in our experimental results, in addition to a lowering of Schottky barrier height. 
\\
In terms of the device aspect, to extract the strain-dependent Schottky barrier height from the two-probe current (Figure 3b(i)), we assume back-to-back Schottky barriers at the drain and source contacts and employ the modified thermionic emission equation (1) shown below\cite{das2013SBHMoS2}. 
\begin{equation}
I_{DS} = SA^*T^{3/2} e^{\frac{-q\phi _B}{\eta _1 k_BT}} [1-e^{\frac{qV_{DS}}{\eta _2k_BT}}]
\end{equation}
\begin{equation}
\implies \ln{\frac{I_{DS}(V_P)}{I_{DS}(V_P= 0 V)}= -\frac{q \Delta \phi _B}{k_BT}}
\end{equation}
Here, $S$ is the area of the device, $A^*$ is the effective Richardson constant (whose value is assumed to be independent of strain)\cite{vabbina2015RicharsonConst}, $T$, $q$ and $k_B$ are the absolute temperature, electronic charge and Boltzmann's constant, respectively. $\eta _1$ and $\eta _2$ are two ideality factors, of which the former was kept fixed at 1 and the latter, which accounts for non-idealities like bias-dependent series and shunt resistances, was varied to obtain a good fit for each $I-V$ trace. In Figure 3b(ii), the thermionic equation-based fit lines are shown overlaid on the corresponding scatter data points for $V_P$= \SI{-3}{}, \SI{0}{}, and \SI{3}{V}. The $\phi _B$ parameter for each $V_P$ trace can be used to calculate the strain-induced change in Schottky barrier height with respect to $V_P$= \SI{0}{V}. The extracted $\phi _B$ values vary from \SI{72}{} to \SI{44}{} to \SI{-108}{meV} as $V_P$ is tuned from \SI{-3}{} to \SI{0}{} to \SI{3}{V}, consistent with reports from literature.\cite{kwon2017thickness} 
\\
Using equation 2, we have also calculated the change in Schottky barrier height ($\Delta \phi _B$) for each trace by considering that no strain is applied at $V_P$= \SI{0}{V}. A plot of $\Delta \phi _B$ with piezo bias $V_P$ (effectively, the strain) extracted from the fits as well as calculated from the ratio of the drain currents is shown in Figure 5b. For compressive strains at negative $V_P$, the $\Delta \phi _B$ is positive, which turns negative and increases in magnitude with increasing tensile strain values for positive $V_P$. Odd-layered \ce{MoS2} flakes have an inherent piezoelectric nature due to the broken inversion symmetry which gives rise to structural anisotropy. The additional strain could lead to a build up of piezoelectric-based polarization charge at the electrodes that would increase (for compressive) or decrease (for tensile) the Schottky barriers. Hence, a piezotronic effect induced modulation of the source Schottky barrier governing electron injection in the \ce{MoS2} channel could also affect the drain current. Representative band diagrams showing the reduction (increase) in bandgap and the corresponding lowering (increase) in Schottky barrier height with tensile (compressive) strain are shown in Figure 5c.
\\
To compare the performance of our all-electrical \ce{MoS2} strain sensor with  literature, we calibrate the voltage-dependent strain in our measurements with typically employed strain values. For this, the reported shift in the position of the $E'$ Raman peak with mechanical strain is compared with the change in peak position with applied piezo voltage in our device. Assuming a similar shift for both kind of strains, we find that for negative $V_P$ the correlated compressive strain per unit negative $V_P$ is in the range of \SI{0.10}{}--\SI{0.14}{\%} per V. \cite{datye2022,yang2014lattice,pak2020strain} For tensile strain per unit positive $V_P$, the correlated strain values are \SI{0.06}{}--\SI{0.08}{\%} per V. Hence, the total applied strain can be varied from \SI{-0.23}{\%} to \SI{0.14}{\%} for $V_P$ ranging from \SI{-3}{V} to \SI{3}{V}. These correlated values can be used to calculate the conventional strain gauge values for our devices (S10, Supplementary Information). In addition, we can also calculate the strain precision that can be achieved through the application of $V_P$ by firstly correlating the $V_P$(strain)-induced change in drain current with a reference strain value (in $\%$) and then evaluating the minimum strain value based on the standard deviation of the drain current (\SI{0.09}{nA} at $V_{DS}$=\SI{-0.05}{V}) measured over repeated $V_P$ switching measurements in Figure 3e. Based on a conservative estimate, we find that a strain precision of 0.002$\%$ is possible by the piezo biasing. Moreover, the Raman-based strain correlation also helps in estimating a strain resolution of 0.045$\%$ for tensile and 0.078$\%$ for compressive strain, respectively, based on the repeated drain current sweep measurements for $V_P$= \SI{0}{} and $V_P$=\SI{1}{V} (Table 2 in Supplementary Information section S11).
\\
Gauge factor values calculated by the change in resistance method for different 2D materials as well as quasi-2D and bulk systems are benchmarked in Figure 5d. It can be seen that 2D materials show significantly larger values of strain gauge in comparison to silicon or metal-based sensors.\cite{fiorillo2018theory, french1989piezoresistance, tsai2015flexible} This is due to the high tensile strength of 2D materials and large density of states at the band edges. Our devices (with a conservative estimate) can show negative (-1498) and positive (1056) gauge factors with significant tunability because of the application of both kinds of strain, which is not typically observed in literature.  2D materials provide the unique advantage of high gauge factors at ultra-thin flake thicknesses.\cite{datye2022, wu2014piezoelectricity, zhu2019vanadium, neri2018electronic, wagner2018highly, smith2013electromechanical} Also, for the case of 2D materials such as \ce{WSe2} and \ce{SnS2}, improved gauge factors have been obtained through ultra-violet illumination\cite{yan2021giant} or reduced temperatures,\cite{qin2021strain} respectively. Nanowire, nanobelt, and microwire structures of \ce{ZnSnO3}, \ce{ZnO}, \ce{SiC}, or \ce{Si} also exhibit considerable gauge factors, however at the cost of significantly larger device thicknesses.\cite{zhou2008flexible, wu2012ultrahigh, pang2012flexible, li2020giant, biccai2019negative} Further, it can be noted that, anisotropic 2D materials, such as \ce{ReS2}, could, in principle, exhibit large gauge factors owing to a stronger dependence of electrical properties on strain.\cite{an2019opposite, zhang2017strain} It should be noted that calculating gauge factors by employing the change in current ($\frac{\Delta I}{I_0}/\Delta \epsilon$), as also seen in literature, can lead to markedly higher gauge factor values (reaching 28000 in our devices).\cite{lee2019ultrahigh, tsai2015flexible, qi2015piezoelectric}
\\
Further, a comprehensive benchmarking table in S11, Supplementary Information compares the different straining strategies specifically employed for \ce{MoS2} and evaluates the qualitative device implementation feasibility of such methods with respect to the substrate used, strain-resolution provided by the technique, precision in the strain-dependent parameters over several cycles, and the ease of technology integration for applications such as CMOS and micro-electro-mechanical systems. It has been highlighted that this work is the only all-electrical study of strain transfer on 2D materials which has been implemented for device applications. 
\\
\section{Conclusion}
In summary, we have demonstrated a novel electro-mechanical device by coupling a piezoelectric thin film on a \ce{Si} substrate with an atomically thin 2D material. Electrically induced strain through the piezo film has been used to tune the physical and electrical properties of 2D monolayer and three-layer \ce{MoS2} over a wide range, and with a high degree of ease, reversibility, precision and resolution. In contrast to typical strain application strategies which involve the use of mechanical apparatus or substrate distortions to apply strain, the device architecture employed in this work enables complete electrical control on a \ce{Si} substrate which can be smoothly integrated with conventional micro-electro-mechanical circuits and systems. Specifically, the shift in characteristic Raman modes, variation in photoluminescence peaks, and the piezoresponse strain loop demonstrate the physical nature of the strain transfer. In addition, the efficient modulation of two and three-terminal electrical characteristics with the piezo voltage-induced strain demonstrates a strain-tuned 2D \ce{MoS2} field effect transistor. It also provides a much needed route to modulate intrinsic electrical properties, such as carrier mobilities, of an as-exfoliated or as-grown 2D material. As a strain sensor, the advantage of controlling the nature of strain by simply using bias polarity helps in realizing large positive (tensile) as well as negative (compressive) gauge factors. Thus, our work provides an interesting and exciting platform for coupling the semiconducting properties of 2D materials and piezoelectric polarization in complex perovskite oxide thin films on silicon substrates which will be useful for future CMOS and MEMS applications. Besides technological relevance for micro-electro-mechanical circuits and systems, it offers a new path for exploring localised strain effects in 2D systems along with exotic strain-derived phenomena such as exciton funnelling and the anistropy-induced bulk photovoltaic effect.\cite{jiang2021flexo, peng2020strain}
    \section{Methods}
    \subsection{Device Fabrication and Characterization}
    The entire device fabrication can be divided into two steps- (i) electrode-piezoelectric film-electrode part and (ii)  the \ce{MoS2} field-effect transistor part. Firstly, selected area bottom electrodes (B) were fabricated by electron-beam lithography (Raith 150-Two) patterning using an EL9/PMMA-950K bilayer resist and metal deposition by sputtering (AJA International) Ti/Pt (5/50 nm), followed by lift-off in acetone. Then, the base layer (LSCO) and the piezoelectric thin film  (PNNZT) were deposited by pulsed laser deposition at \SI{600}{\degreeCelsius} and \SI{750}{\degreeCelsius}, respectively at an oxygen partial pressure of \SI{0.13}{mbar}.\cite{ade2018} A KrF excimer laser at a repetition rate of \SI{5}{Hz} and energy density of \SI{1.5}{J/cm^2} was used for the pulsed deposition from high-quality LSCO and PNNZT targets prepared by solid state reaction method. Post-deposition, the substrate was allowed to cool down slowly in oxygen environment. The top electrodes (T) were then e-beam lithography patterned by carefully aligning over the bottom electrodes, followed by Ti/Pt (3/25 nm) deposition and lift-off, such that an MPM stack can be formed.
    \\
    To electrically separate the MPM stack from the 2D material, a thin dielectric film of \SI{15}{nm} \ce{Al2O3} was deposited by plasma atomic layer deposition at \SI{200}{\degreeCelsius}, at a growth rate of 1 \AA/cycle. Thereafter, an ultra-thin flake of \ce{MoS2} was micro-mechanically exfoliated and transferred on top of the MPM region. Source and drain electrodes on \ce{MoS2} were fabricated by consecutive electron-beam lithography, metal deposition of Ni/Au (20/80 nm) and lift-off processes. The piezoelectric response of PNNZT was obtained using piezoresponse force microscopy performed on an MFP-3D AFM (Asylum Research) using Dual AC Resonance Tracking (DART) mode. X-ray diffraction patterns were recorded using a PANalytical Empyrean system with with Cu K$_{\alpha}$ ($\lambda$ = 1.54 \AA) radiation. Raman and photoluminescence spectroscopies was carried out on a LabRAM HR Evolution (HORIBA Scientific) with 532 nm laser, 700 nm spot diameter, using gratings with 1800 and 600 groves/mm, respectively. All the electrical characterization were carried out on a probe station connected to a Keysight B1500A parameter analyzer equipped with multiple source-measure units (SMUs).
    
    \subsection{Density Functional Theory Calculations}
    All computations were implemented using Vienna Ab initio Simulation Package (VASP) and the generalized gradient approximation of Perdew–Burke–Ernzerhof (PBE) was used to describe the exchange correlation.\cite{kresse1996efficient, kresse1999ultrasoft} The projector augmented wave (PAW) pseudopotentials were used for the frozen core-valence electron interactions and the dispersion corrected van der Waals DFT-D3 method was used.\cite{grimme2010consistent} For both uniaxial and biaxial strained structures, a plane wave cut-off of 500 eV with $\Gamma$-centered \textit{k}-points grid of $15\times15\times1$ were used for geometry optimization. For the self-consistent calculations, an energy cut-off of 500 eV and denser \textit{k}-points grid of $21\times21\times1$ were used. All optimizations were continued until an energy convergence criteria of 0.1 meV was reached and the Hellman–Feynman forces were less than 0.01 eV \AA$^{-1}$. Vacuum of \SI{20}{\AA} was incorporated in the transverse direction to avoid spurious interactions across adjacent periodic images.

\begin{acknowledgement}
The authors thank Prof. Udayan Ganguly, Dr. Kartikey Thakar and Himani Jawa for critical discussions. The authors acknowledge the Indian Institute of Technology Bombay Nanofabrication Facility (IITBNF) for usage of its device fabrication and characterization facilities. S.L. acknowledges support from Department of Science and Technology, Government of India through its SwarnaJayanti fellowship scheme (Grant No. DST/SJF/ETA-01/2016-17). The authors acknowledge computational support from the Australian National Computing Infrastructure (NCI) and Pawsey supercomputing facility for high performance computing. Y.Y. and N.V.M. acknowledge the support from the Australian Research Council (CE170100039). 

\end{acknowledgement}


%% file: SI.tex
\newpage 
\section{S1: Growth of Thin Films }
Pulsed laser deposition technique was used for the growth of the piezoelectric and buffer layers. A \ce{KrF} excimer laser of wavelength \SI{248}{nm} at a repetition rate of \SI{5}{Hz} was incident on single phase targets of PNNZT, LSCO, or STO. The key parameters used for pulsed laser deposition of STO, LSCO and, PNNZT thin films are shown in the table below. The spot size of the laser was \SI{1.5}{mm}$\times$\SI{3.5}{mm}.

\begin{table}[h]
    \centering
    \begin{tabular}{||c c c c c c||}
\hline
Material & Energy density & No. of shots & Partial Pressure \ce{O2} & Temp. & Thickness  \\[0.5ex] 
& (J/cm$^2$) & & (mbar) & ($^O$C) & (nm) \\ [0.5ex] 
\hline\hline
STO & 1.5 & 500 $-$ 1000 &  0.13 & 750 & 50 $-$ 100 \\
LSCO & 1.5 & 500 $-$ 1000 & 0.13 & 600 & 50 $-$ 100 \\
PNNZT & 1.3 & 4000 $-$ 6000 & 0.13 & 750 & 350 $-$ 500 \\  [1ex] 
 \hline
    \end{tabular}
    \caption{Summary of key process parameters used for pulsed laser deposition of given thin films}
    \label{tab:my_label}
\end{table}
\section{S2: Properties of PNNZT film}
PNNZT thin films were deposited on a platinized \ce{SiO2}/\ce{Si} substrate, from single phase ceramic targets which were prepared via solid state reactions at high temperatures. XRD pattern in the figure below shows the major diffraction planes of the as-prepared target.
\begin{figure}[H]
  \includegraphics[scale=0.7]{PNNZT_target_XRD.png}
   \caption{XRD patterns of the target used for PNNZT deposition.}
\label{sch:FigureS41}   
\end{figure}
PNNZT films were deposited on STO or LSCO buffer layers to prevent the formation of the pyrochlore phase which can deteriorate the piezoelectric properties of the film. The use of a buffer layer enables epitaxial growth of the PNNZT thin film. Scanning electron micrograph was used to obtained surface-level topographical information of the film.
\begin{figure}[H]
\includegraphics[scale=1.05]{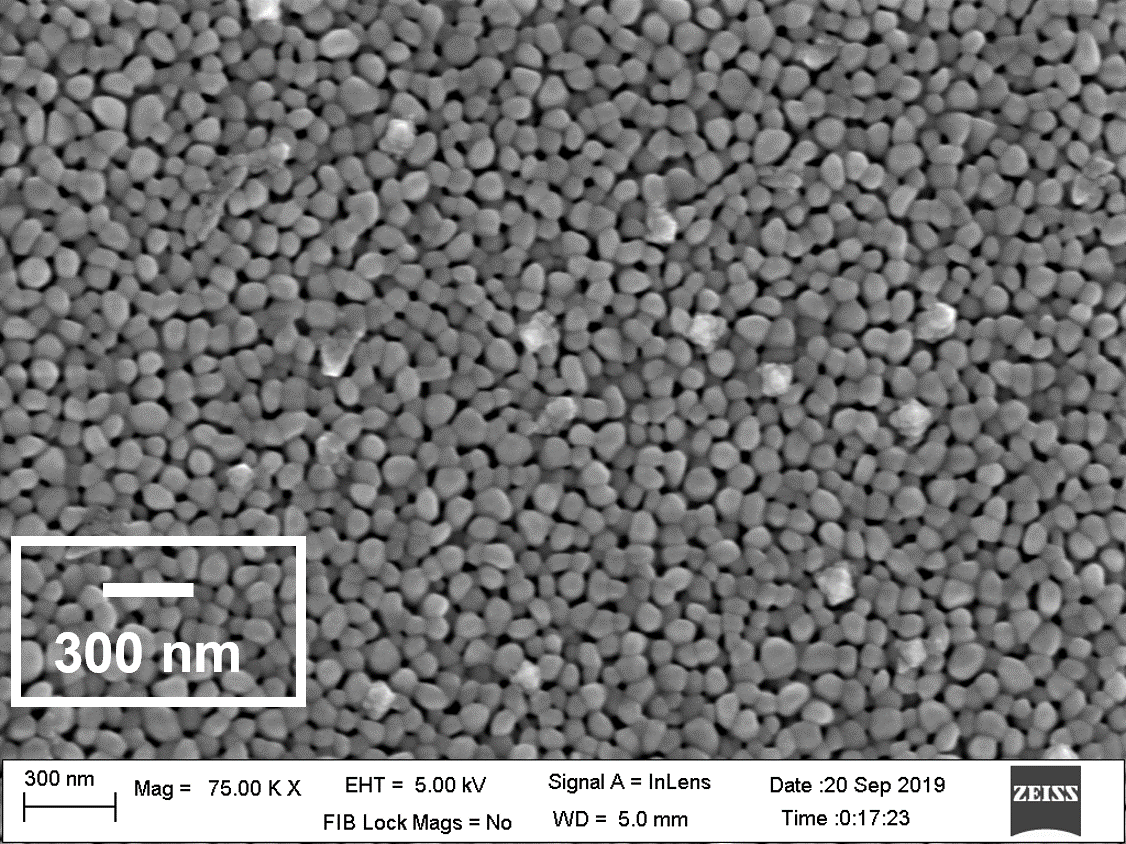}
   \caption{ SEM image of the deposited PNNZT thin film.}
\label{sch:FigureS2221112}   
\end{figure}
PNNZT is a relaxor ferroelectric material which shows large values of piezoelectric coefficients, $d_{33}$ and $d_{31}$. PFM studies were used to image the domains in PNNZT and examine the phase switching of these ferroelectric domains. 
\begin{figure}[H]
\includegraphics[width=\textwidth,height=\textheight,keepaspectratio]{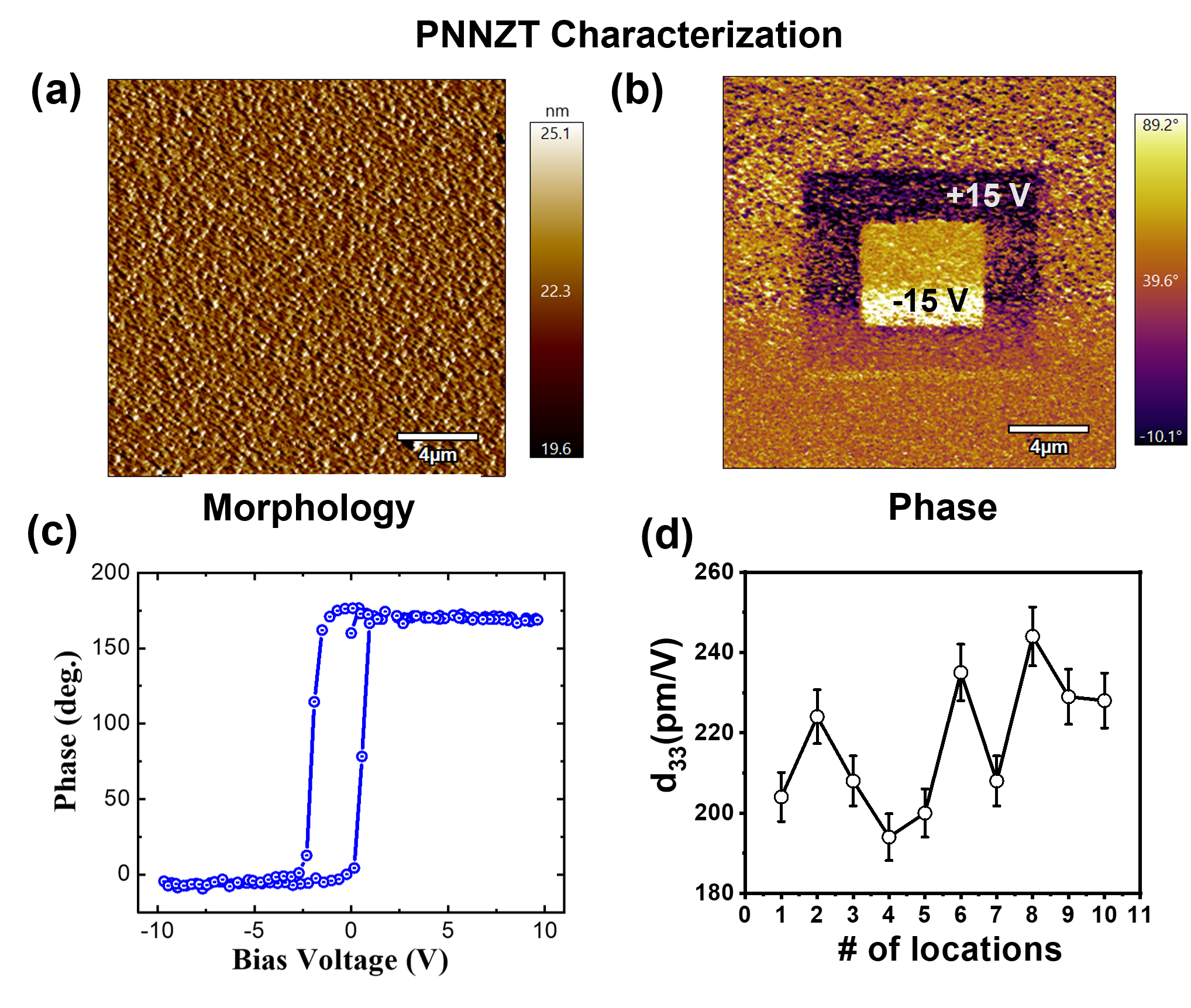}
   \caption{\textbf{a}, \textbf{b} PFM images depicting the morphology and phase of PNNZT thin film, respectively. \textbf{c} Phase switching and \textbf{d} $d_{33}$ values obtained from various positions on the film from the hysteresis strain loop under DC biasing.}
\label{sch:FigureS44}   
\end{figure}
In response to the vertical electric field applied by the tip, the strain hysteresis loop can be obtained. A linear fit of the displacement vs. bias voltage gives the $d_{33}$ component of the PNNZT film. 
Further, the polarization vs. electric field (P-E) loop of the film shows a large remnant polarization of 29.5 $\mu$Ccm$^{-2}$.
\begin{figure}[H]
  \includegraphics[scale=0.4]{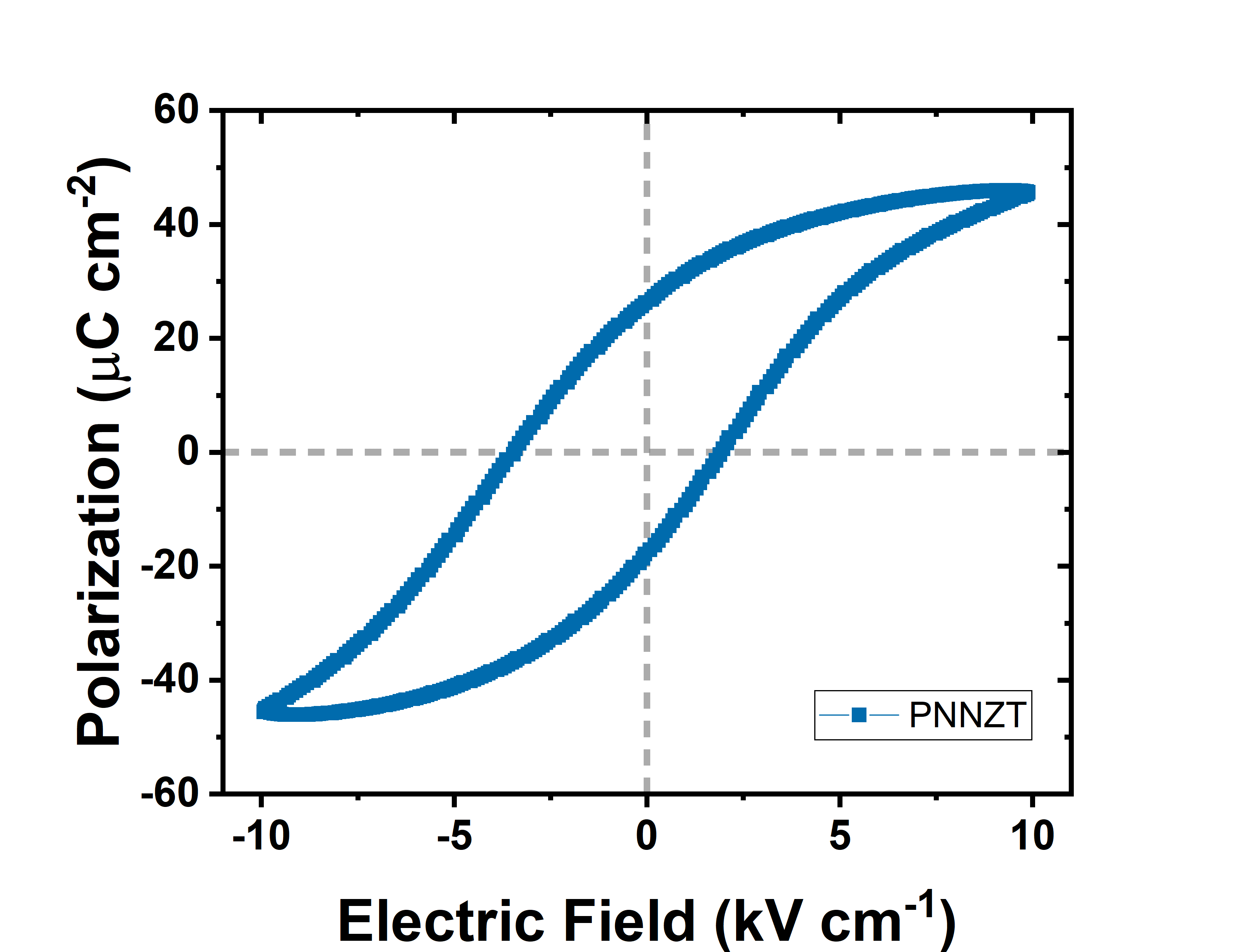}
   \caption{P-E hysteresis loop of the PNNZT film. }
\label{sch:FigureS3}   
\end{figure}
\section{S3: Atomic force microscope image of \ce{MoS2}}
AFM imaging (\ce{MoS2} flake region marked by dotted lines) was carried out for thickness identification and surface roughness analysis. The thickness of \ce{MoS2} flake (along marked line) is found to be around \SI{2.4}{nm} implying that the flake is three-layered. The mean surface roughness is around \SI{0.9}{nm}.
\\
\begin{figure}[H]
  \includegraphics[scale=0.7]{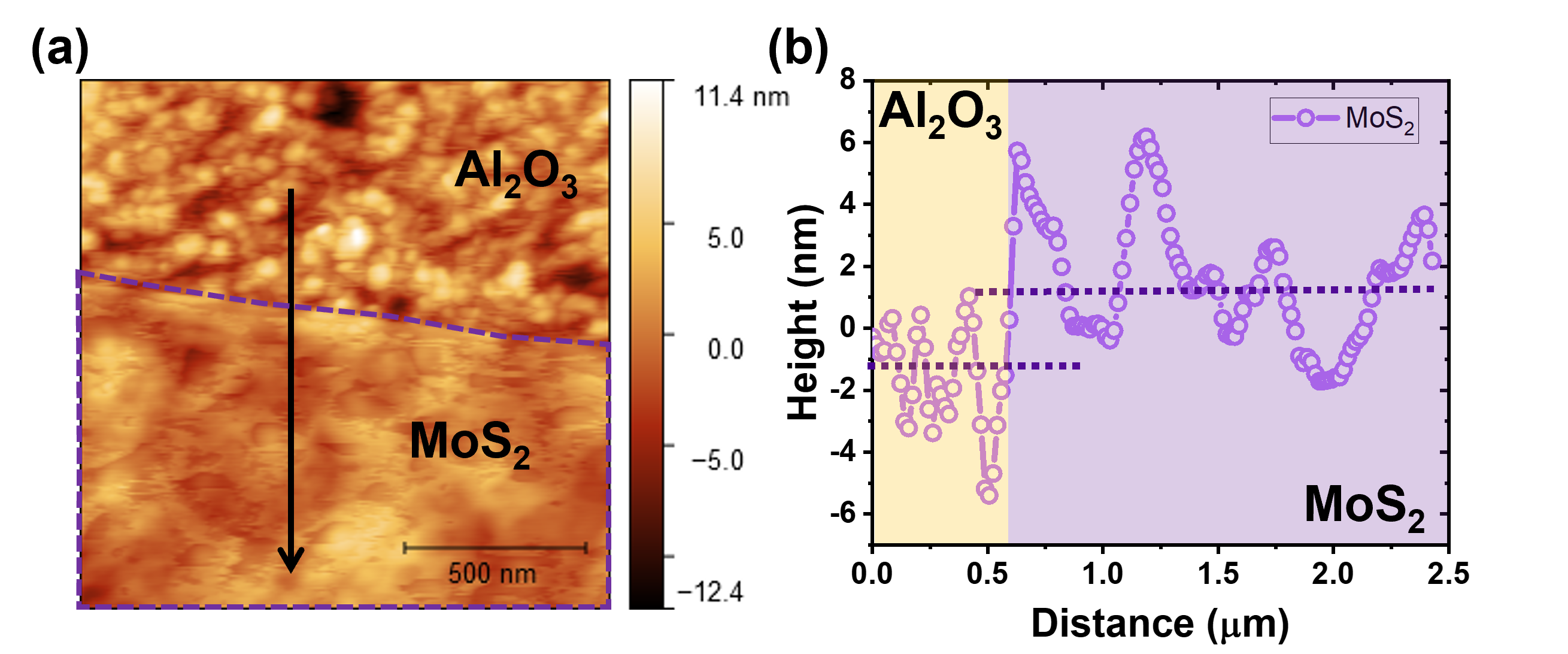}
   \caption{AFM image of \ce{MoS2} flake on the \ce{Al2O3} surface. }
\label{sch:FigureS55}   
\end{figure}


\section{S4: Density Functional Theory based Bandstructure Calculations}
\subsubsection{Methodology}
The DFT simulations were implemented using Vienna Ab initio Simulation Package (VASP) using the generalized gradient approximation of Perdew–Burke–Ernzerhof to describe the exchange correlation. The projector augmented wave (PAW) pseudopotentials were utilised to account for the frozen core-valence electrons.\cite{kresse1996efficient, kresse1999ultrasoft} van der Waals interaction is incorporated by the dispersion corrected DFT-D3 method.\cite{grimme2010consistent} For structural optimization, a plane wave cut-off of 500 eV and a $\Gamma$-centered \textit{k}-points grid of 15$\times$15$\times$1 were used. For the self-consistent calculations, an energy cut-off of 500 eV and a denser \textit{k}-points grid of 21$\times$21$\times$1 were used. All optimizations were performed until the energy converged to a limit of 0.1 meV and the Hellman–Feynman forces  were less than 0.01 eV$/$\AA. Sufficient vacuum of at least 20 \AA was incorporated in all structures in the vertical direction to avoid any spurious interactions between periodic images.
\subsubsection{Bandstructure of 3L-\ce{MoS2} under uniaxial strain}
We applied uniaxial strain along X-direction and allowed the unit cell to relax in the Y-direction so that the effect of strain was redistributed during the structural relaxation. The bandstructures of the unstrained, tensile and compressive strained structures, respectively, are shown in Figure S6 below. Under tensile strain, the conduction band minimum is at \textbf{K} point, while for compressive strain, it shifts to the \textbf{Q'} point along \textbf{K} $-$ \textbf{$\Gamma$}. Thus, the effective mass of electrons (in n-type \ce{MoS2}) is calculated for these two valleys.
\begin{figure}
\includegraphics[width=\textwidth,height=\textheight,keepaspectratio]{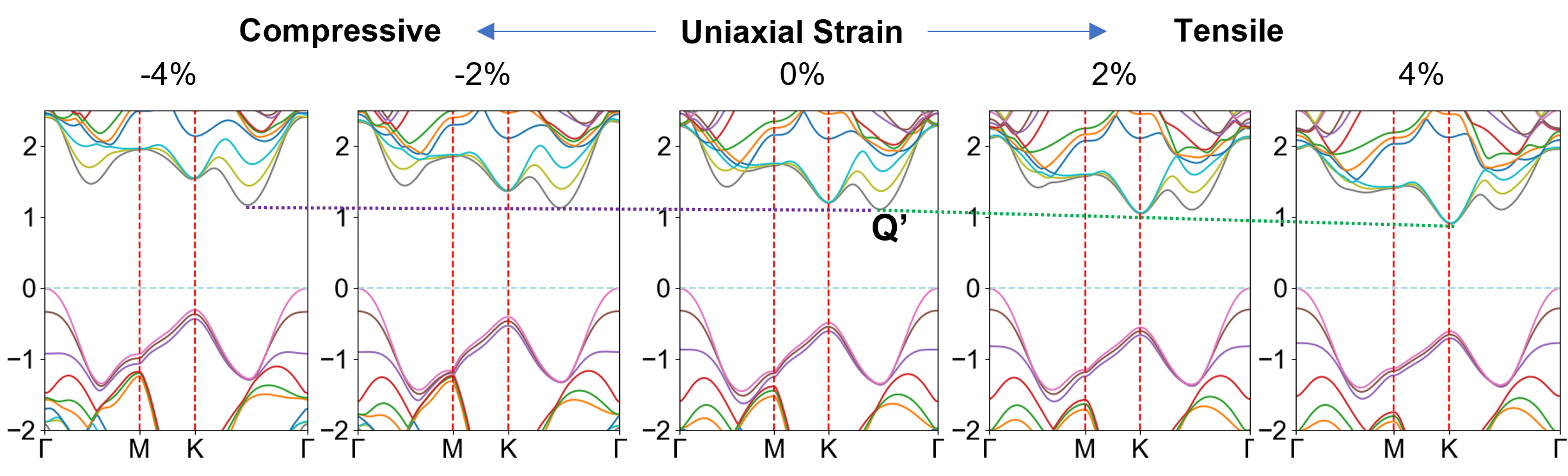}
   \caption{Evolution of electronic bandstructure of 3L-\ce{MoS2} when uniaxially strained from compressive to tensile regimes.}
\label{sch:FigureS42}   
\end{figure}

\subsubsection{Bandstructure of 3L-\ce{MoS2} under biaxial strain}
For biaxial strain, the lattice was stretched or compressed uniformly along both X and Y directions. The change in bandgap and effective mass values is more significant in comparison to uniaxial strain.
\begin{figure}[H]
\includegraphics[width=\textwidth,height=\textheight,keepaspectratio]{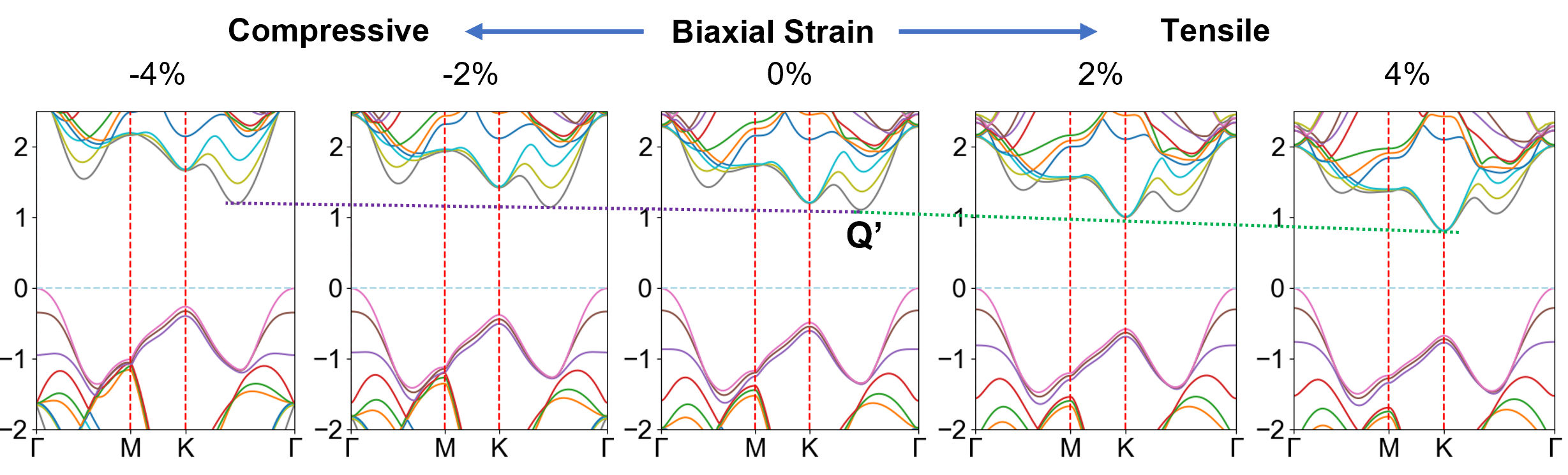}
   \caption{Evolution of electronic bandstructure of 3L-\ce{MoS2} when biaxially strained from compressive to tensile regimes. }
\label{sch:FigureS34432}   
\end{figure}

\subsection{Effective mass from bandstructures}
For the strained structures, the effective mass of electrons was calculated at the respective conduction band minima. The effective mass decreases for both tensile and compressive strains. The effective masses calculated at the \textbf{Q'} valley for tensile strains show a slight increase, however, the energy difference between the valley at \textbf{Q'} and the corresponding conduction band mimimum (at \textbf{K}) is significantly large, hence they are represented by near-transparent symbols.
\begin{figure}[H]
\includegraphics[scale=1]{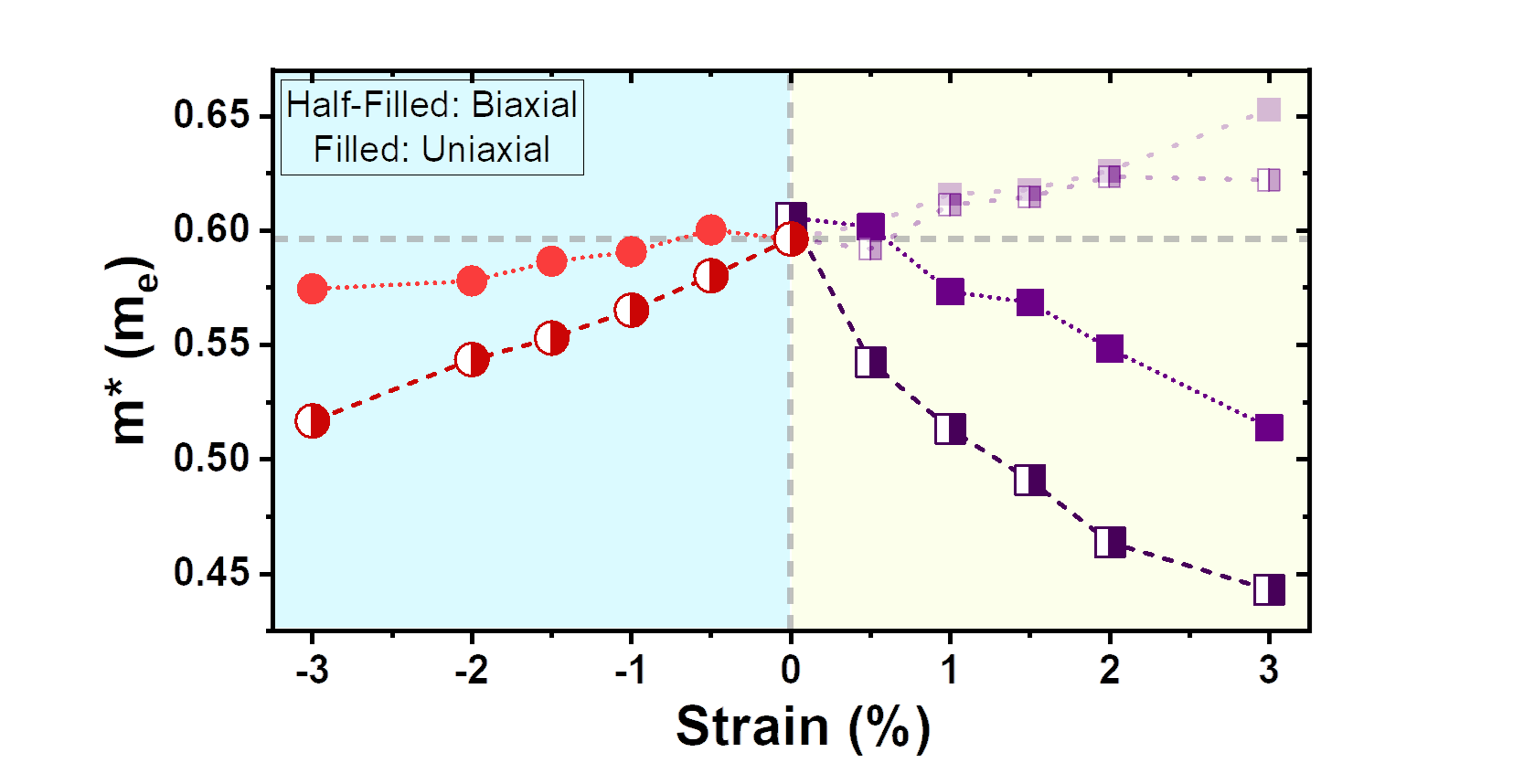}
   \caption{Change in effective mass of electrons with tensile and compressive strains of uniaxial and biaxial nature. The effective mass of electrons at the respective conduction band minima decreases for both tensile and compressive strains. The effective masses calculated at the \textbf{Q’} valley for tensile strains show a slight increase, however, the energy difference between these valleys and the corresponding conduction band mimimum (at \textbf{K}) is significantly large, hence they are represented by near-transparent symbols.}
\label{sch:FigureS53332}   
\end{figure}

\section{S5: PFM analysis of \ce{MoS2} on non-piezo substrate}
The PFM strain hysteresis loops of \ce{MoS2} flakes on non-piezoelecric substrates like \ce{Si}\slash \ce{SiO2} are shown in the figure below. The displacements of the PFM tip due to any piezoelectric effect from the surface of \ce{MoS2} are recorded. 
\begin{figure}[H]
  \includegraphics[width=\textwidth,height=\textheight,keepaspectratio]{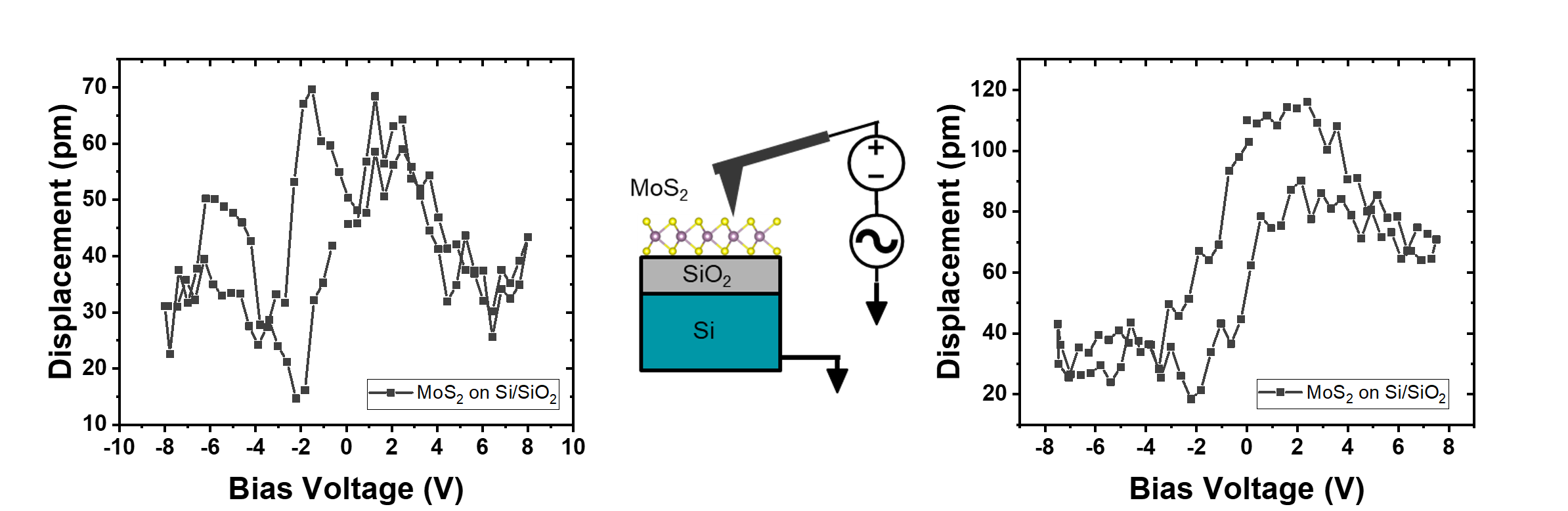}
   \caption{Strain hysteresis loops from PFM scans of \ce{MoS2} on \ce{Si}\slash \ce{SiO2} surface. A schematic of the experimental configuration is also shown.}
\label{sch:FigureS213}   
\end{figure}
\section{S6: Strain-tuned I-V: Additional Devices}
The modulation of two-probe electrical transport of two additional devices is shown in the figure below. The trace of unstrained ($V_P$= \SI{0}{V}) is shown as black dotted points.
\begin{figure}[H]
\includegraphics[width=\textwidth,height=\textheight,keepaspectratio]{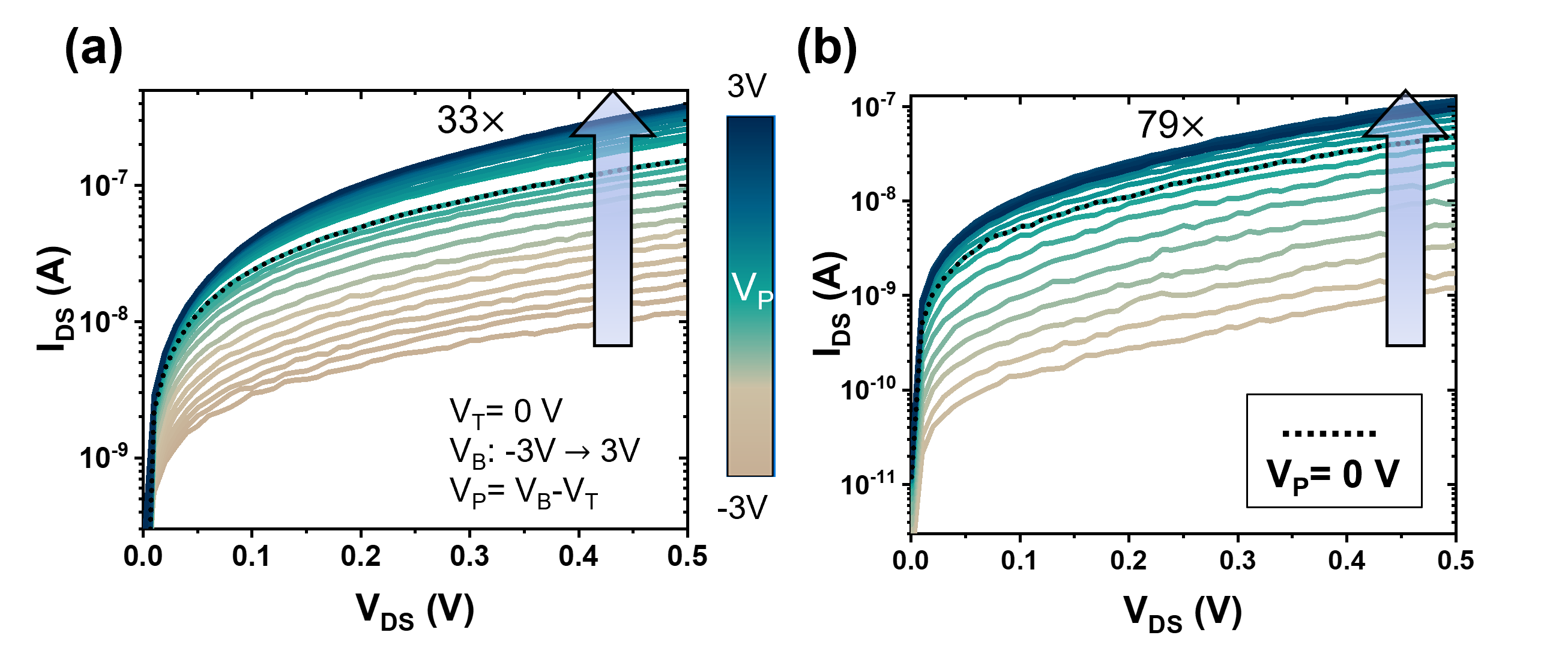}
   \caption{\textbf{a}, \textbf{b} Two-probe transport modulated by strain in two different devices.}
\label{sch:FigureS5555552}   
\end{figure}

\section{S7: Multiple Sweeps of Output Electrical Characteristics for Different Strain Conditions}
It is important to ascertain the variation in drain current over multiple output characteristic sweeps, to understand the strain resolution of the device. The plot of drain current with its standard deviation at different $V_{DS}$ values for $V_P$= \SI{0}{V} (unstrained) and $V_P$= \SI{1}{V} (tensile strained) conditions is shown. Over ten sweep cycles, the standard deviation in the current for both cases is an order of magnitude lower than the corresponding drain current values at any specific $V_{DS}$. The clear difference in the mean $I_{DS}$ values indicates that the \ce{MoS2} transistor can unambiguously resolve strain difference corresponding to a change of \SI{1}{V} in $V_P$. 
\begin{figure}[H]
\includegraphics[scale=1.10]{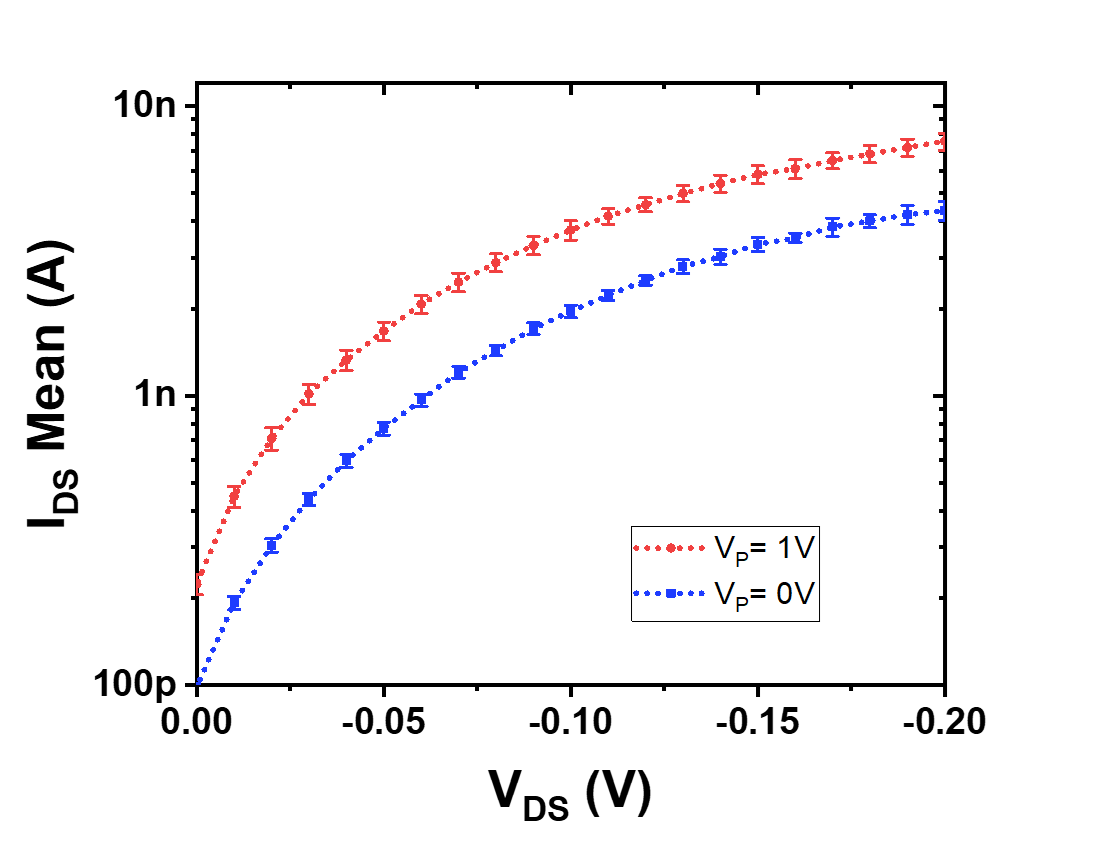}
   \caption{Output characteristics showing the mean current and standard deviation in the unstrained and strained cases over ten sweeps for each $V_P$.}
\label{sch:FigureS2222}   
\end{figure}

\section{S8: Reversibility of Transfer Characteristics}
The \ce{MoS2} FET transfer characteristics were acquired before, during, and after the offset-bias measurements described in Figure 4. This is to check the reversibility of the strain-modulated transistor characteristics. The corresponding plots are overlaid for reference.
\begin{figure}[H]
\includegraphics[scale=1.10]{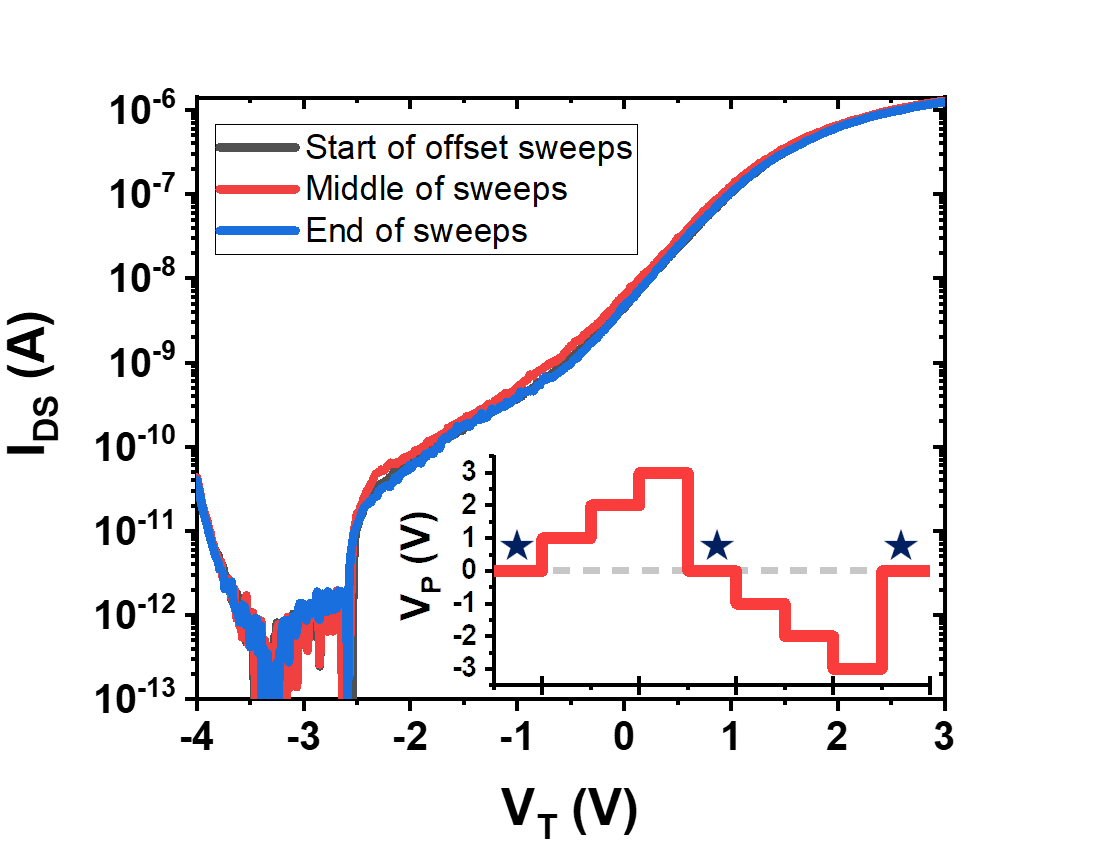}
   \caption{Overlaid transfer characteristics under \SI{0}{V} offset bias value at different instances of the varying strain measurement cycle marked with stars.}
\label{sch:FigureS12122}   
\end{figure}

\section{S9: Comparing Strain-effect with Field-effect for \ce{MoS2} transistor performance}
To evaluate the contributions of strain-effect and field-effect in our devices, we employ different biasing configurations on the top and bottom electrodes: 
\begin{enumerate}
\item $V_T$ = $V_B$ = varying: Top and bottom electrodes are shorted so that there is no electric field across the piezoelectric layer. However, the \ce{MoS2} channel will be subjected to a gate field due to the voltage on the top electrode. Hence, the field-effect will be the only mechanism for charge carrier modulation in the channel. As seen in the $I_{DS}-V_{DS}$ characteristics below, the current in the \ce{MoS2} channel can be driven from $on$ state ($V_T$ = \SI{2}{V}) to $off$ state ($V_T$ = \SI{-2}{V}) with significant modulation of drain current between these two points. It can be seen from Figure 4 of the main manuscript that the threshold voltage of the \ce{MoS2} FET is around \SI{1}{V}.
\item $V_T$ = varying, $V_B$ = \SI{0}{V}: Here, the biasing is opposite in nature to that employed for evaluating the strain-effect in Figure 3 of the main manuscript. The top electrode is biased from \SI{-2}{V} to \SI{2}{V} which can switch the transistor from $off$ to $on$ due to the field-effect. However, since $V_B$ is set at \SI{0}{V}, there is an electric field across the piezoelectric film which can lead to a strain-effect. This electric field is opposite in direction to the typically applied fields in Figure 3b in the main manuscript. There are some key observations to be noted here. Firstly, under high positive $V_T$, the $I_{DS}$ values are similar to those obtained for case (i) (red curves) implying that field-effect is dominant in this regime. Secondly, it can be seen that for negative $V_T$, when the transistor is in a high resistance state, its drain current values are higher than the $off$ currents in (i) (compare blue and violet curves in (i) and (ii)). This is because the channel is in the $off$ state and is not flooded by charge carriers ($V_T$ $<$ $V_{Th}$), hence, strain-effect is more prominent. The direction of strain electric field is such that a tensile strain is applied on the \ce{MoS2} flake which, as seen in Figure 3b, can result in an increased drain current. Thus, this biasing scheme leads to a strain-effect based modulation of field-effect limited drain current, especially through tensile strain at sub-$V_{Th}$ gate biases.
\item $V_T$= \SI{0}{V}, $V_B$ = varying: This is the typical biasing scheme employed in obtaining the strain-dependent traces in Figure 3b of the main manuscript. The top electrode voltage is set at \SI{0}{V} which is below the threshold voltage of the transistor. Hence, the transistor is in the $off$ state and just the strain-effect can be used to modulate the drain current. 
\end{enumerate}
\begin{figure}[H]
\includegraphics[width=\textwidth,height=\textheight,keepaspectratio]{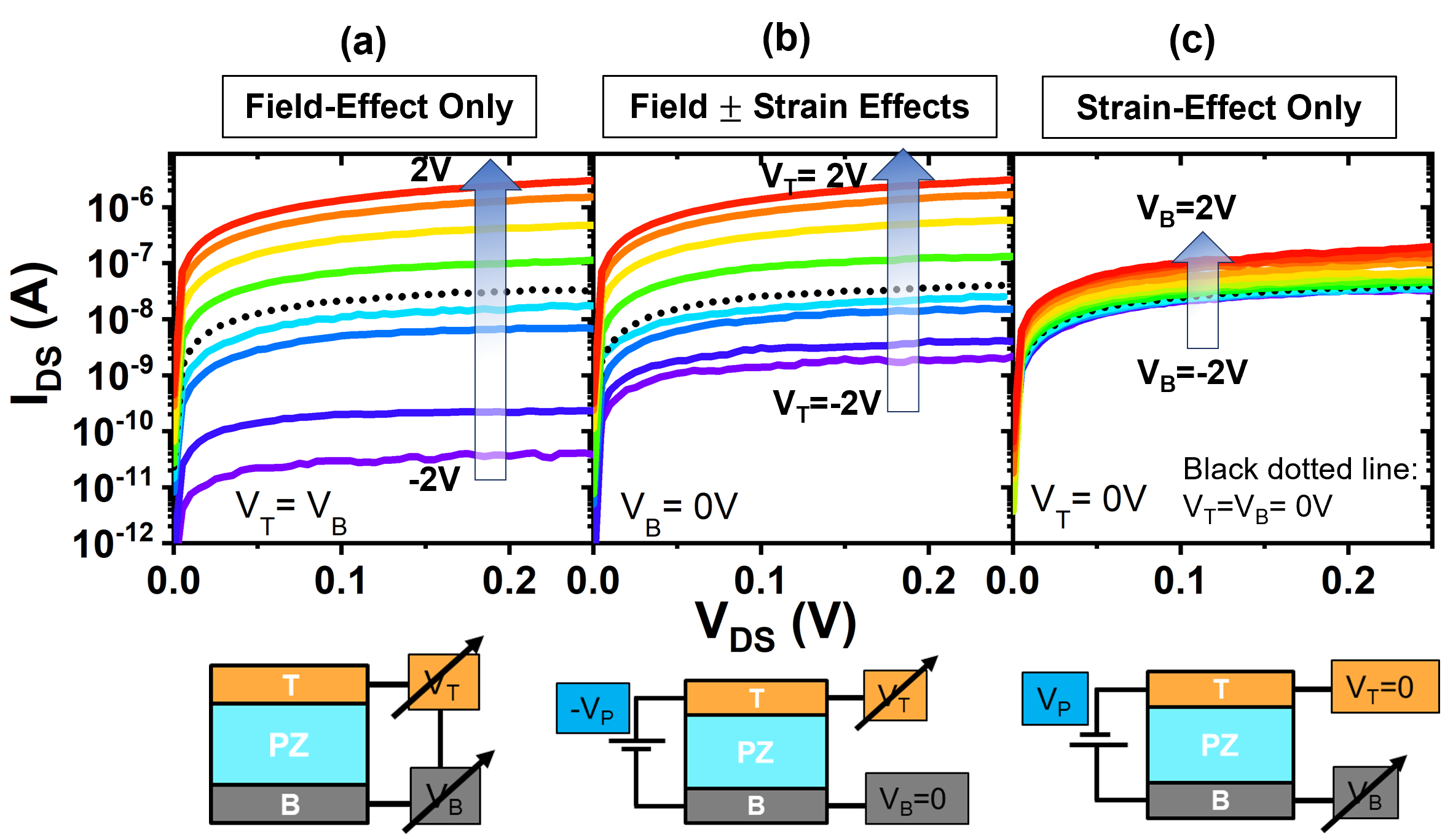}
   \caption{Comparison of $I_{DS}-V_{DS}$ characteristics under three different biasing schemes which are used to segregate the strain-effect and the field-effect in the \ce{MoS2} transistor. }
\label{sch:FigureS2212}   
\end{figure}

\section{S10: Strain Gauge Calculation}
It is important to have a quantitative estimate of the amount of strain transferred from the piezoelectric thin film to the 
 \ce{MoS2} layer in terms of commonly reported values in literature. Strain-dependent Raman peak shifts can provide a correlation between the nature of strain as well as the magnitude of strain. Hence, by comparing the Raman shifts in this work with literature reports, we can quantify the strain transferred. It should be noted that, in literature, the exact strain on 2D material flakes is not typically calculated, however, the $\%$ values that are reported are specific to the mechanical set-ups employed for the straining experiments. 
\\
To calculate the \textit{conventional} strain gauge values, the Raman peak shifts from our work were calibrated using published reports.\cite{datye2022, pak2017strain, yang2014lattice} Thereafter, the \textit{shift per volt} in our piezoelectric straining method is calculated from the \textit{shift per strain} as shown in the figure below.

\begin{figure}[H]
\includegraphics[width=\textwidth,height=\textheight,keepaspectratio]{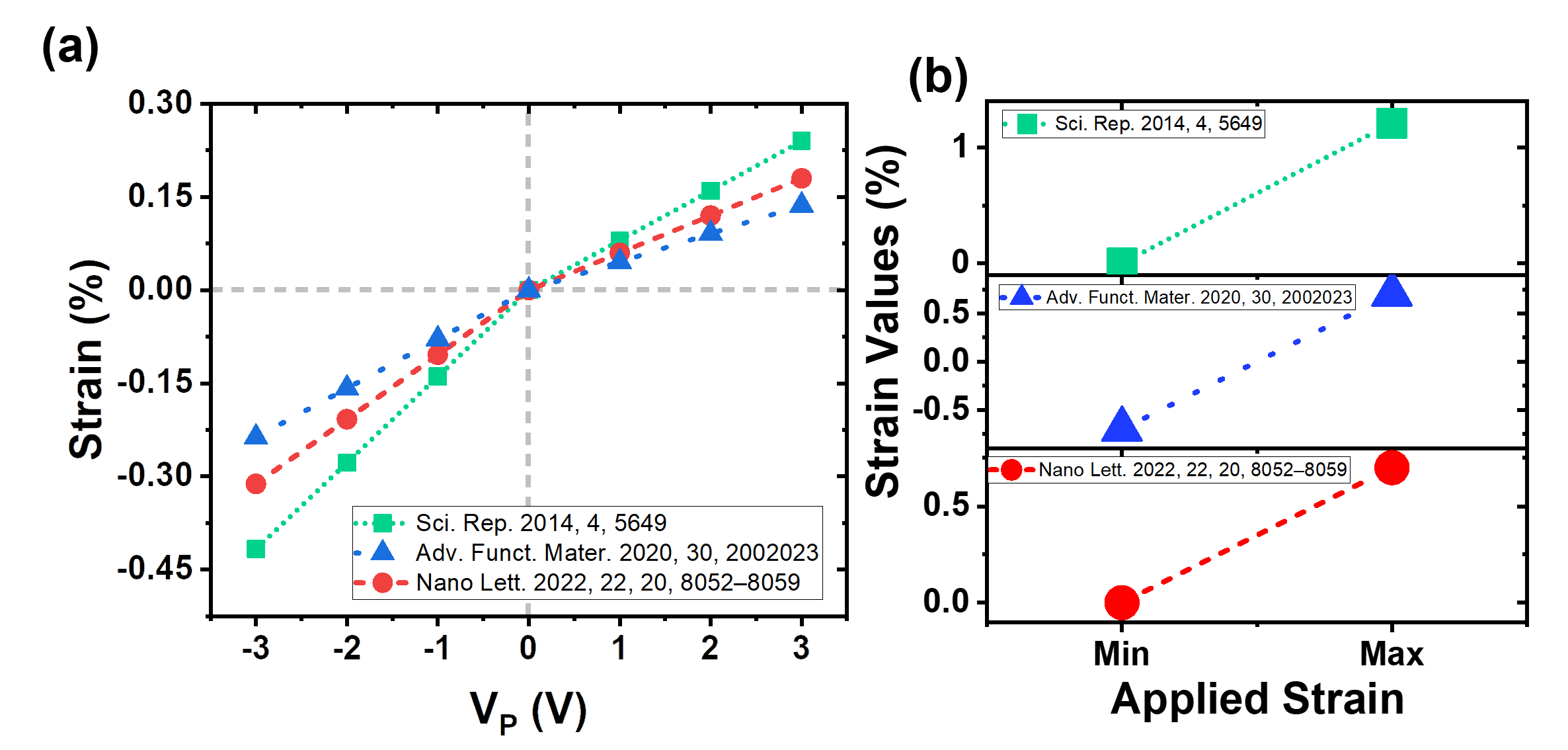}
   \caption{\textbf{a} Correlation between strain values obtained from mechanical strain-transfer techniques and our electrical piezo voltage-based technique obtained by mapping Raman peak shifts in both cases. \textbf{b} Typical range of strain values applied using mechanical set-ups in literature.}
\label{sch:FigureS1112222}   
\end{figure}

Based on these correlations, the \textit{conventional} strain gauge values were calculated. A comparison of \textit{conventional} strain gauge values for our device calculated from strain-dependent Raman shifts of three published reports is shown in the Figure below. 
\begin{figure}[H]
\includegraphics[width=\textwidth,height=\textheight,keepaspectratio]{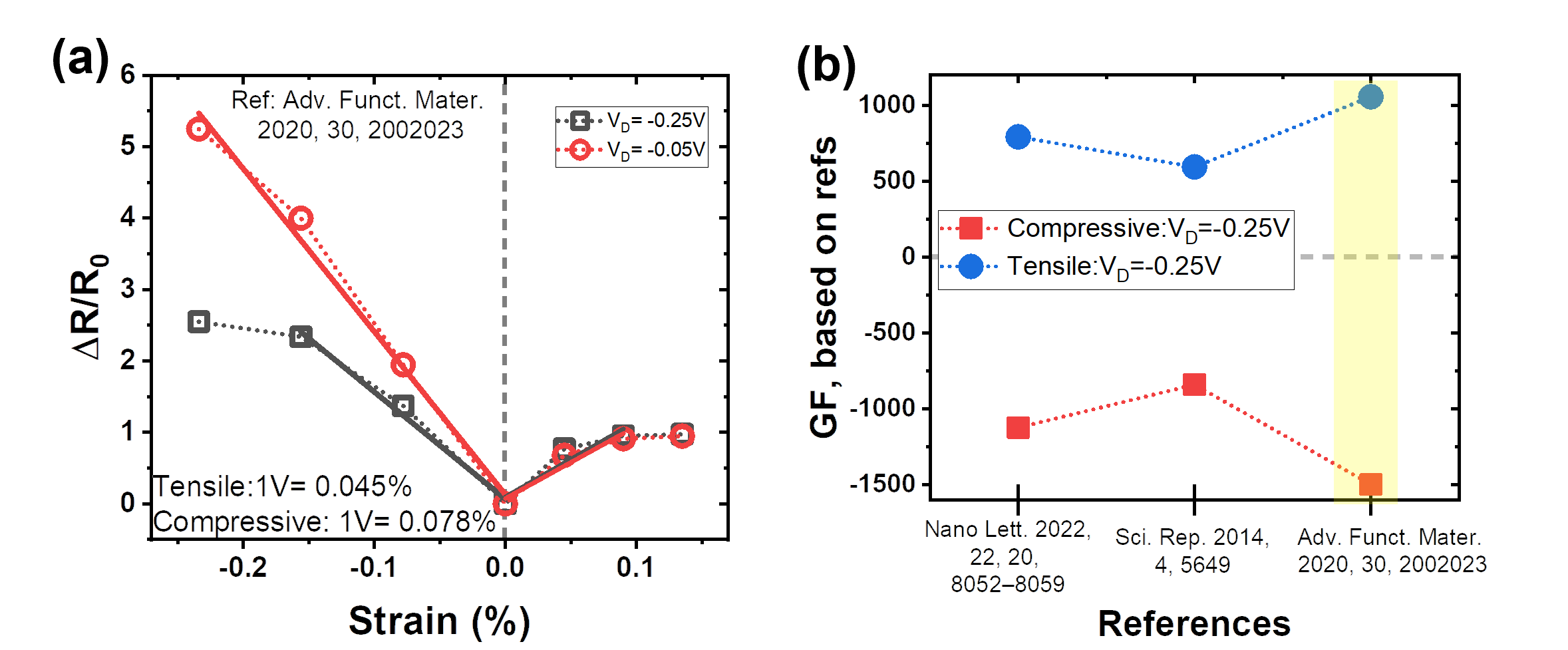}
   \caption{\textbf{a} Estimation of gauge factor from a normalised change in resistance vs. calibrated strain plot. The strain is calibrated based on Adv. Funct. Mater. 2020, 30, 2002023. \textbf{b} Comparison of gauge factor values calculated based on strain-dependent Raman shifts from the provided references.}
\label{sch:FigureS11212112}   
\end{figure}

\section{Section S11: Benchmarking table}
\begin{table}[h!]
    \centering
     \begin{adjustbox}{width=\textwidth}
    \begin{tabular}{|c| c| c| c| c| c| c| c| c| c| c|}
\hline
Ref&Method/Substrate & Transduction & \multicolumn{2}{c|}{\makecell{Nature \\ of Strain}}  & Resolution & Precision & Ease & \multicolumn{2}{c|}{Gauge Factor} & \makecell{CMOS/MEMS\\ Integration} \\ 
\hline
  &  &  &  T & C & & & & T & C &\\
 \hline
 \cite{datye2022} & Flexible Substrate- PEN & Mech$\rightarrow$ Mech & \checkmark & \ding{53}  & Low & Low & Easy & 150 & - & Difficult\\
 \hline
  \cite{pak2017strain} & Flexible Substrate- PET & Mech$\rightarrow$ Mech & \checkmark & \checkmark & Low & Low & Easy & - & - & Difficult\\
  \hline
  \cite{wu2014piezoelectricity} & Flexible Substrate- PET & Mech$\rightarrow$ Mech & \checkmark & \ding{53} & Low & Low & Easy & 760$^\ast$ & - & Difficult\\
 \hline
   \cite{lee2019ultrahigh} & Flexible Substrate- PET & Mech$\rightarrow$ Mech & \checkmark & \checkmark & Low & Low & Easy & 575294$^\ast$ & - & Difficult\\
 \hline
 \cite{tsai2015flexible} & Flexible Substrate- PET & Mech$\rightarrow$ Mech & \checkmark & \ding{53} & Low & Low & Easy & 40$^\ast$ & - & Difficult\\
  \hline
  \cite{qi2015piezoelectric} & AFM tip & Mech$\rightarrow$ Mech & \checkmark & \checkmark & High & High & Difficult & 1160$^\ast$ & - & Difficult\\
 \hline
  \cite{manzeli2015piezoresistivity} & AFM tip & Mech$\rightarrow$ Mech & \ding{53} & \checkmark & High & High & Difficult & - & \SI{-148}{} & Difficult\\
 \hline
\cite{ng2022improving} & Pre-patterned Substrates & Structural$\rightarrow$ Mech & \checkmark & \ding{53} & Low & Low & Difficult & - & - & Difficult\\
\hline
\cite{jaikissoon2022mobility} & Capping layers & Structural$\rightarrow$ Mech & \checkmark & \ding{53} & Low & Low & Difficult & - & - & Possible\\
\hline
\cite{hui2013} & Piezo substrate & Elec$\rightarrow$ Mech & \ding{53} & \checkmark & High & High & Easy & - & - & Possible \\
\hline
Our & \makecell{Piezo thin film \\ on Si substrate} & Elec$\rightarrow$ Mech & \checkmark & \checkmark &  \makecell{T=0.045$\%$ \\ C=0.078$\%$ \\ (High)} &  \makecell{0.002$\%$ \\ (High)}  & Easy & 1056 & \SI{-1498}{} & Easy \\
\hline

    \end{tabular}
    \end{adjustbox}
    \caption{A comparison of different straining methods employed for \ce{MoS2} in literature, in terms of their ease of implementation and performance metrics. \\
    Resolution is the ability to resolve a change in device parameters (like, drain current) with the smallest applied strain. Precision accounts for achieving a similar value of a parameter over multiple cycles. Ease corresponds to a qualitative ease of device implementation based on different straining techniques. Tech integration refers to the utilization of the strain transfer methods in standard CMOS-based devices and micro-electro-mechanical circuits. T and C represent tensile and compressive strains, respectively. $^\ast$ mark denotes gauge factor value calculated using ($\frac{\Delta I}{I_0}$) instead of the change in resistance method.}
    \label{tab:my_label}
\end{table}

\bibliography{achemso-demo}